%% file: main.tex
\documentclass{article}
\usepackage[preprint]{neurips_2025}

\usepackage{amsmath, amssymb, amsfonts}

\usepackage{graphicx}
\usepackage{booktabs}
\usepackage{multirow}
\usepackage{caption}
\usepackage{subcaption}
\usepackage[section]{placeins}

\usepackage{hyperref}
\usepackage{url}

\usepackage{microtype}
\usepackage{xcolor}

\graphicspath{{figs/}{./figs/}{../figs/}}

\hypersetup{
  colorlinks=true,
  linkcolor=blue!50!black,
  citecolor=blue!50!black,
  urlcolor=blue!50!black,
  pdfborder={0 0 0}
}

\providecommand{\code}[1]{\texttt{#1}}
\providecommand{\pkg}[1]{\textbf{#1}}

\title{CANNs: A Toolkit for Research on Continuous Attractor Neural Networks
}

\author{%
  Sichao He$^{1,\dagger}$ \And Aiersi Tuerhong$^{1, 2,\dagger}$ \And Shangjun She$^{1}$ \AND
  Tianhao Chu$^{1}$ \And Yuling Wu$^{1}$ \And Junfeng Zuo$^{1}$ \And Si Wu$^{1,\ast}$
  \AND
  \\[-\smallskipamount]
  $^{1}$School of Psychological and Cognitive Sciences, Peking--Tsinghua Center for Life Sciences,\\
  PKU-IDG/McGovern Institute for Brain Research, Center of Quantitative Biology,\\
  Peking University, Beijing, China
  \\
  $^{2}$College of Mathematics and Statistics, Chongqing University, Chongqing, China
  \\
  $^{\dagger}$These authors contributed equally.
  \\
  $^{\ast}$Corresponding author: \texttt{siwu@pku.edu.cn}
}

\date{}

\begin{document}
\maketitle

\begin{abstract}
Continuous attractor neural networks (CANNs) are the canonical
computational framework for how the brain encodes continuous variables
such as spatial position, head direction, and movement direction, and
explain the activity of hippocampal place cells, entorhinal grid cells,
and head-direction cells. CANN research, however, is fragmented: most
results rest on lab-specific implementations, general-purpose
simulators lack CANN-specific abstractions, and the path from spike
trains to attractor geometry in real recordings lacks a standardized
toolkit. Here, we present a comprehensive open-source toolkit that
unifies the full CANN research workflow. It combines three tightly
integrated components: 1)~\textbf{canns}, a Python library on
BrainPy/JAX that provides standardized 1D/2D CANNs,
spike-frequency-adaptation variants, grid cell networks, hierarchical
path-integration models, and brain-inspired attractor architectures,
together with curated datasets, task generators, an analyzer module
and trainer modules for biologically plausible plasticity;
2)~\textbf{canns-lib}, a Rust acceleration backend delivering
hundreds-of-times speedups for spatial-navigation workloads and modest
gains for Ripser-based persistent homology; 3)~\textbf{ASA} (Attractor Structure Analyzer), a PySide6
pipeline applying persistent homology and cohomology to experimental
neural recordings to detect ring-like and toroidal attractor
signatures in real data. The toolkit ships with full-detail
reproducible pipelines that recover recent CANN results including
SFA-driven anticipative tracking, theta sweeps in
head-direction/place/grid systems, and hierarchical path integration.
\end{abstract}

\input{sections/intro}
\input{sections/toolkit}
\input{sections/asa}
\input{sections/results}
\input{sections/backmatter}

\bibliographystyle{plainnat}
\bibliography{refs}

\end{document}

%% file: sections/intro.tex
%

\section{Introduction}

\subsection{Background: CANN as a canonical neural circuit model}

How does the brain represent continuous variables such as spatial
position, head direction, and movement direction, and update them in
real time as an animal moves through its environment? These questions
lie at the heart of systems neuroscience. A remarkably productive
answer, accumulated over half a century, is the \emph{continuous
attractor neural network} (CANN). A CANN is a recurrently connected
population of neurons whose collective activity forms a localized
``bump'' that can be smoothly translated along a low-dimensional
manifold. The bump's position encodes the variable of interest; the
attractor structure endows the representation with stability, noise
robustness, and integration properties that a feedforward code cannot
match.

The CANN framework was developed by several authors in the history, such as the seminal work of \citet{amari1977dynamics}, which 
developed the lateral-inhibition type neural fields in the context of dynamics theory. There have been a large number of variants of CANN models, in particular, \citet{wu2008dynamics} and
\citet{fung2010moving,wu2016continuous} proposed an analytically solvable CANN model, which greatly facilitates the research on CANNs. The biological case for CANNs is grounded
in three landmark discoveries: hippocampal place cells
\citep{o1971hippocampus}, which fire at specific spatial locations;
entorhinal grid cells \citep{hafting2005microstructure}, whose
periodic hexagonal firing fields tile space; and head-direction cells
\citep{taube1990head}, which encode angular orientation. The
attractor perspective has since been extended to spike-frequency
adaptation \citep{mi2014spike,li2025dynamics}, theta rhythmicity
\citep{chu2024firing,ji2025phase,ji2025systems}, hierarchical path
integration \citep{chu2025localized}, and toroidal population
geometry in grid cells \citep{Gardner2022Toroidal}. CANNs are now
widely regarded as a canonical model of neural information
representation---analogous in status to the Hodgkin--Huxley model at
the single-neuron level.

The CANN models in this library are primarily based on the
mathematically tractable and canonical continuous attractor neural
network called the Wu--Amari--Wong (WAW) model
\citep{wu2008dynamics,fung2010moving,wu2016continuous}.
This canonical model provides an elegant theoretical framework for
understanding continuous attractor dynamics, and its mathematical
tractability enables researchers to deeply analyse network stability,
dynamical properties, and encoding capabilities.

\subsection{The fragmentation problem}

Theoretically rich as CANNs are, the field suffers from a practical
problem: \emph{fragmentation}. Most published results rest on
lab-specific implementations; a graduate student wishing to reproduce
a figure from a recent paper typically has to rebuild the model from
scratch, often without access to the original code. The
re-implementation cost is non-trivial because CANN dynamics are
sensitive to integration schemes, parameterizations, and stimulus
representations, and small deviations can qualitatively change
bump-tracking behaviour. Three concrete consequences follow.

First, the entry barrier is high. Newcomers to CANN research must
spend weeks implementing lateral-inhibition dynamics, path-integration
loops, and SFA-modulated equations before they can run a single
simulation. Second, published results are hard to compare. A reported
phase-precession slope or anticipative-tracking lead time cannot be
evaluated against a model that uses a different connectivity
parameterization or a different numerical integrator. Third, the path
from experimental data to attractor geometry is essentially
unstandardized: there is no widely adopted toolchain for taking spike
trains from, for example, a grid-cell recording and quantifying the
torus-like structure of the population code.

General-purpose neural simulators---NEST \citep{Gewaltig:NEST},
Brian~2 \citep{stimberg2019brian}, and NEURON \citep{hines1997neuron}---excel
at large-scale spiking-network simulation but provide no
CANN-specific abstractions. Domain-specialized code in research labs
does not generalize, is rarely versioned for release, and typically
lacks analysis tooling. In short, CANN research today resembles
pre-Transformers NLP: every group reinvents the wheel.

\subsection{Why a comprehensive toolkit}

The natural remedy is a comprehensive, well-engineered, openly
released toolkit that provides a uniform API, validated
implementations, and reproducible workflows. The CANNs ecosystem
described in this paper is our attempt at such a remedy. The
comprehensive scope is deliberate: a library that covers only
simulation is useful for modelling but not for data analysis; a
toolkit that covers only data analysis is disconnected from theory.
By packaging the Python library (\texttt{canns}), the Rust
acceleration backend (\texttt{canns-lib}), and the Attractor
Structure Analyzer (ASA) pipeline together, we aim to support the
full cycle---from mathematical model, to simulation, to analysis of
experimental recordings---under a consistent set of abstractions. The
target audience is threefold: computational neuroscientists who build
and validate CANN models; experimental neuroscientists who want to
test for attractor signatures in their recordings; and students and
newcomers who want to learn CANN concepts through runnable,
documented examples.

\subsection{Contributions}

The contributions of this paper form a complete ecosystem for CANN
research, organised around three tightly integrated components
(the \texttt{canns} library, the \texttt{canns-lib} Rust backend, and
the ASA pipeline) together with reproducible baselines and documentation
materials.

\begin{enumerate}
  \item \textbf{The \texttt{canns} Python library.} A modular
        five-layer architecture (Application / Functional / Core
        Models / Foundation / Hardware) built on BrainPy/JAX
        \citep{wang2023brainpy,jax2018github,wang2025integrating} that
        provides standardized implementations of 1D/2D CANNs, SFA
        variants \citep{mi2014spike,li2025dynamics}, grid cell
        networks \citep{burak2009accurate}, hierarchical path
        integration models \citep{chu2025localized}, theta-sweep
        models \citep{chu2024firing,ji2025phase,ji2025systems}, and
        brain-inspired attractor architectures
        \citep{amari1977neural,hopfield1982neural}; curated datasets;
        task generators; an analyzer module (model metrics, data
        analysis, slow-point / fixed-point analysis, and visualization);
        and a trainer module for biologically plausible plasticity
        rules.
  \item \textbf{The \texttt{canns-lib} Rust acceleration backend.}
        A native companion library that delivers substantial
        speedups for the two heaviest computational kernels in the
        ecosystem: persistent homology via Ripser
        \citep{carlsson2009topology,edelsbrunner2010computational}
        (1.13$\times$ on average, up to 1.82$\times$) and
        long-trajectory spatial navigation ($\sim$700$\times$
        vs.\ the pure-Python reference; see \S3.2).  The
        backend is invoked transparently from the Python API.
  \item \textbf{The ASA pipeline.} A \pkg{PySide6}-fronted
        application that takes experimental spike trains or dense
        population activity through embedding, cell
        classification, point-cloud construction, persistent
        homology and cohomology, and circular-coordinate decoding,
        producing interpretable ring and toroidal
        attractor signatures---validated on real grid-cell
        recordings and consistent with the toroidal topology
        reported by \citet{Gardner2022Toroidal}.
  \item \textbf{Reproducible re-implementations of recent SOTA
        results.}  Full-detail runnable pipelines that
        \emph{re-implement} the following recent results using the
        toolkit, rather than merely citing them: SFA-driven
        anticipative tracking \citep{mi2014spike,li2025dynamics};
        theta sweeps in head-direction, grid, and place cell
        systems \citep{chu2024firing,ji2025phase,ji2025systems}; and
        localized/phase coding in hierarchical grid-cell networks
        \citep{chu2025localized}.  In each case the corresponding
        author of the original work is also a co-author of this
        report; the re-implementations are intended as
        reproducibility baselines, not as independent confirmations
        of the original results.
  \item \textbf{Documentation, tests, and bilingual support.}
        Bilingual (English/Chinese) Sphinx documentation, a
        continuous-integration test suite, and a curated gallery
        of worked examples.
\end{enumerate}

\section{Background and Related Work}
\label{sec:background}

\subsection{Continuous attractor neural networks}

The continuous attractor neural network has a compact mathematical
core. In a widely studied form---the Wu--Amari--Wong (WAW) model
\citep{wu2008dynamics,fung2010moving,wu2016continuous}---
$N$ neurons are arranged on a 1D ring or 2D torus, the recurrent
connectivity follows a translation-invariant profile normalized by divisive normalization, and
the population dynamics admit a continuum of stable bump states
parameterized by a continuous variable. The bump can be moved
arbitrarily by a localized external input, and the network's
attractor structure provides a built-in integrator: when the input
disappears, the bump remains at its last position, enabling
short-term memory. These properties are derived analytically in the
ring case and extend naturally to 2D and higher-dimensional
manifolds.

CANNs have been applied to three canonical neural systems. Head
direction \citep{taube1990head} is naturally modeled as a 1D ring
attractor. Spatial position in the hippocampus
\citep{o1971hippocampus} requires 2D place-field representations and
is closely linked to grid cells \citep{hafting2005microstructure}
in the entorhinal cortex, whose periodic hexagonal firing fields
arise naturally in attractor network models with path-integration
dynamics \citep{samsonovich1997path,burak2009accurate,
mcnaughton2006path}. Recent work has extended CANNs in two
biologically important directions. First, spike-frequency adaptation
(SFA) introduces an activity-dependent negative feedback that turns
the bump into a moving wave and supports anticipative tracking
\citep{mi2014spike,li2025dynamics}. Second, theta-rhythmic
modulation produces theta sweeps, phase precession, and procession
in head-direction, grid, and place cell networks
\citep{chu2024firing,ji2025phase,ji2025systems,wang2015theta}. A
more recent development combines localized and phase coding into
hierarchical grid-cell networks \citep{chu2025localized}.

\subsection{Topological data analysis for neural activity}

A second line of work relevant to this report is topological data
analysis (TDA) for neural population activity. The foundational
machinery is persistent homology
\citep{carlsson2009topology,edelsbrunner2010computational}, which
summarizes a point cloud as a multiscale barcode of topological
features. For neural applications, persistent homology on the
time-embedded point cloud of population activity is used to detect
1-dimensional loops (ring attractors) and 2-dimensional voids
(toroidal attractors), as in \citet{Kim2017RingAttractor} for
ring-attractor circuits and \citet{Gardner2022Toroidal} for grid
cell population geometry.

Persistent cohomology and the related duality machinery
\citep{Vaupel2023Duality} extend the approach by computing
cohomology classes (cocycles) directly, from which smooth circular
coordinates can be extracted and mapped back onto the underlying
behaviour. The ASA pipeline (Section~4) builds on this machinery:
it takes either a time-embedded spike-train point cloud or a
dense activity matrix, computes barcodes, performs a shuffle
control for significance, decodes cohomology circular coordinates,
and projects them back into physical space as a CohoMap or onto
the animal's trajectory as a CohoSpace. A complementary perspective
uses fixed-point and slow-manifold analysis
\citep{sussillo2013opening,golub2018fixedpointfinder} to study the
intrinsic dynamics of trained recurrent networks; the \texttt{canns}
analyzer module includes tools for both approaches.

\subsection{General neural simulators}

The closest existing software to a general CANN platform is the
family of neural simulators that originated in computational
neuroscience. NEST \citep{Gewaltig:NEST} focuses on point-neuron
network simulation at large scale; Brian~2
\citep{stimberg2019brian} provides a flexible Python front-end with
code-generation backends; NEURON \citep{hines1997neuron} is the
standard tool for multi-compartment biophysical neuron models. These
simulators are mature, well tested, and highly efficient for
large-scale spiking networks, but they are agnostic to the
mathematical structure of CANN dynamics: there is no built-in notion
of a translation-invariant connectivity, a bump state, or a
continuous attractor manifold. The user must re-implement CANN
dynamics from primitive elements for every project, and the analysis
tooling for inspecting the resulting activity is general rather than
CANN-specific.

More recently, BrainPy
\citep{wang2023brainpy,wang2025integrating} has emerged as a
JAX-based dynamics programming framework that retains the
expressiveness of a Python simulator while inheriting JIT
compilation and GPU/TPU support from JAX \citep{jax2018github}.
The \texttt{canns} library is built directly on top of BrainPy and
inherits these performance properties, while adding the
CANN-specific abstractions (bump state, translation-invariant
connectivity, attractor-aware analyzers) that the general-purpose
simulators lack.

\subsection{Tools landscape and our position}

Compared with the broader scientific-Python ecosystem, the
\texttt{canns} ecosystem occupies a niche analogous to
Hugging Face Transformers for natural language processing: a
domain-specialized, opinionated library that provides standardized
implementations of canonical models, a uniform task API, and a
shared set of analyses and reproducibility pipelines, all built on
top of a general-purpose framework (PyTorch / TensorFlow for
Transformers; BrainPy / JAX for CANNs).

\paragraph{Neighbouring toolboxes in the CANN / spatial-navigation
space.}
Three communities work on adjacent problems and are worth
distinguishing from \texttt{canns}.  \textbf{RatInABox}
(\url{https://github.com/RatInABox-- rodent-navigation-box}) is
the de-facto reference implementation of 1D / 2D rodent-style
spatial-navigation agents with biologically grounded dynamics
(heading, speed, boundary effects, thigmotaxis).  \texttt{canns}
adopts RatInABox's trajectory representation and reproduces
its environment / agent API in \texttt{canns.spatial}, but
extends it with a CANN head-direction read-out and with a
\textbf{$\sim$700$\times$ Rust / PyO3 acceleration} of the
inner simulation loop on long trajectories (see
\S3.2).  Pure-Python RatInABox scales unfavourably past
$\sim$$10^4$ steps; the \pkg{canns-lib} spatial backend
scales linearly to $10^6$ steps in under a second on a laptop.
\textbf{Brian~2} \citep{stimberg2019brian} and the original
NEST / NEURON simulators are general-purpose spiking-network
frameworks; they can in principle simulate CANN dynamics but
require the user to assemble bump attractors from primitive
elements and to write the analysis tooling from scratch.
\textbf{DMT} (Data Management Toolkit for neuroscience) and
\textbf{NWB} (Neurodata Without Borders) solve orthogonal
problems (data ingestion, format standardisation, provenance
tracking); \texttt{canns} is meant to consume NWB / DMT-formatted
recordings rather than to replace them.

\paragraph{Neighbouring toolboxes in the TDA / persistent-homology
space.}
The reference Ripser implementation is the
Cython/\texttt{C++} package \texttt{ripser.py}
(\url{https://github.com/Ripser/ripser}); it computes the same
Vietoris--Rips persistent barcodes as \pkg{canns-lib}'s
\texttt{ripser} module, and is the baseline we benchmark
against (1.13$\times$ geometric mean, 1.82$\times$ peak).  Pure-
Python \texttt{giotto} and \texttt{gudhi} provide richer TDA
primitives (filtrations beyond Vietoris--Rips, persistent
homology with coefficient rings, zigzag persistence) but at
substantially higher per-call cost; the \pkg{canns-lib} design
choice is to keep the Ripser path tight and call the upstream
libraries only when their advanced features are required.  On
the analysis side, \texttt{elephant} (electrophysiology
analysis) and \texttt{NeuralEnsemble} provide complementary
single-cell and population-coding statistics, but neither
includes a topological-analysis layer for population manifolds.
\texttt{canns} focuses on the topology layer and leaves the
other statistics to those packages.

\paragraph{Position of the CANNs ecosystem.}
Within computational neuroscience, the closest analogues to
\texttt{canns} are Brian~2's auditory-cochlea and spatially
structured network examples, and specific toolboxes for
fixed-point analysis \citep{golub2018fixedpointfinder}, but
none of these provide end-to-end CANN modelling, performance
acceleration, and a TDA-based analysis pipeline for experimental
recordings under a single coherent API.

Concretely, the contributions of this report relative to existing
tools are: (i)~CANN-specific model and task abstractions that
general-purpose simulators do not provide; (ii)~an integrated
analysis suite that combines model metrics, data analysis,
slow-point / fixed-point analysis, and visualization rather than
scattering them across separate toolboxes; (iii)~an optional Rust backend
(\texttt{canns-lib}) that delivers hundreds-of-times speedups for
spatial-navigation workloads and more modest gains
(1.13$\times$ on average, 1.82$\times$ peak) for Ripser-based
persistent homology, without changing the Python interface; and
(iv)~a graphical front end (the ASA pipeline) that brings
TDA-based attractor-structure inference to experimentalists
who do not want to write Python code. The following sections
flesh out each component in turn.

%% file: sections/toolkit.tex

\section{The CANNs Ecosystem: Three Co-Designed Components}

\begin{itemize}
  \item \textbf{\pkg{canns}} (Python): the modeling and analysis
        library.  Implements CANN models, task generators, analysis
        tools, brain-inspired trainers, and pipeline orchestration
        on top of \pkg{BrainPy}/\pkg{JAX}.  This is the layer
        readers will use most often.
  \item \textbf{\pkg{canns-lib}} (Rust): the acceleration backend.
        A standalone Rust crate that provides a drop-in
        performance-critical subset of the toolkit (Ripser-based
        persistent homology, long spatial navigation trajectories,
        bulk task generation) through a thin Python FFI.  Released
        and versioned independently.
  \item \textbf{ASA} (Python + GUI): the experimental-data
        pipeline.  Wraps \pkg{canns} with persistent homology,
        persistent cohomology, and circular-coordinate decoding,
        and exposes the workflow through a \pkg{PySide6} graphical
        front end.  Detailed methods are deferred to \S4; here we
        give the entry points and the role it plays in the
        ecosystem.
\end{itemize}

The three are developed in lockstep but live in separate
repositories, and a user can install any one of them without
pulling in the others.  \pkg{canns} declares \pkg{canns-lib} as
a \emph{hard dependency}, however, so that the Ripser and
spatial-navigation acceleration paths are
always available to anyone who installs \pkg{canns}.  ASA
additionally requires \pkg{canns} as a runtime dependency, and
ships its GUI as the optional \texttt{canns[gui]} extra.

\subsection{\pkg{canns} (Python): the Modeling and Analysis Layer}

The \pkg{canns} library is the component readers use most.  It is
built on two design principles that govern
its module decomposition, public API, and extension interface.

\paragraph{Separation of concerns.}
The library is partitioned into five top-level modules:
\textbf{Models} (\texttt{canns.models}) define neural network
dynamics and an \texttt{update()} step; \textbf{Tasks}
(\texttt{canns.task}) generate experimental paradigms and input
tensors; \textbf{Analyzers} (\texttt{canns.analyzer}) visualize
and analyze simulation outputs \emph{or} experimental recordings
without modifying model state; \textbf{Trainers}
(\texttt{canns.trainer}) implement biologically plausible learning
rules on model parameters; and \textbf{Pipeline}
(\texttt{canns.pipeline}) orchestrates the full cycle
(configuration $\to$ execution $\to$ analysis $\to$ reporting).
Because each module is functionally narrow, the same model can be
coupled to different tasks, the same analyzer can consume outputs
from any model that exposes standard array shapes, and the same
task can drive multiple models.

\paragraph{Extensibility through base classes.}
Every major component inherits from one of three abstract base
classes that pin down the public interface: \code{BasicModel} (in
\texttt{canns.models.basic}) for continuous attractor networks
with fixed connectivity; \code{BrainInspiredModel} (in
\texttt{canns.models.brain\_inspired}) for networks whose weights
are updated by local plasticity rules; and \code{Trainer} (in
\texttt{canns.trainer}) for the learning algorithms that operate
on the models above.  Subclasses implement a small set of
methods (\texttt{init\_state}, \texttt{update}, \texttt{energy}
for energy-based models, \texttt{weight\_attr} for the trainer
contract), which guarantees that custom implementations drop in
alongside the built-in ones.

\subsubsection{Five-Layer Architecture}

\begin{figure}[ht]
  \centering
  \includegraphics[width=\columnwidth]{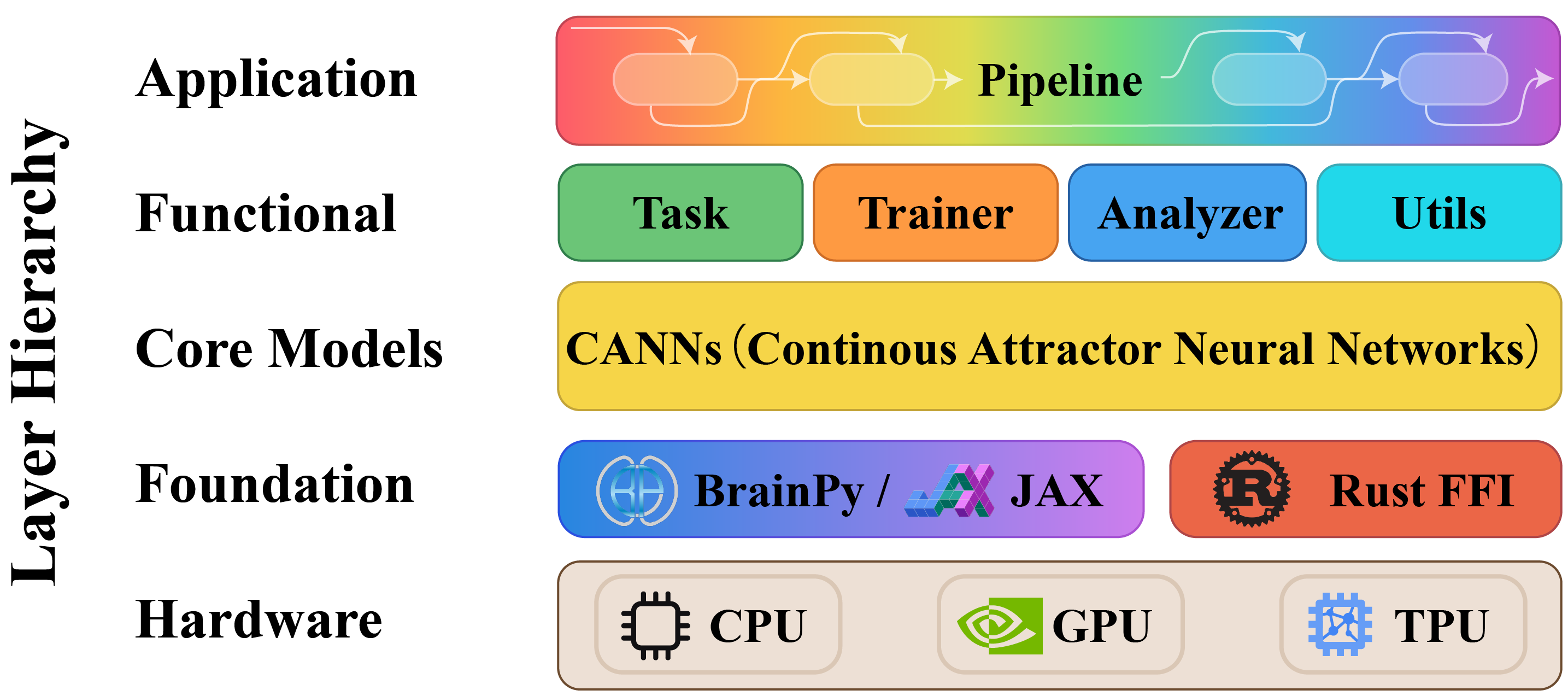}
  \caption{Layer hierarchy of the CANNs ecosystem, showing the five
    layers from hardware up to the user-facing application layer.
    The \textbf{Application} layer exposes pipeline entry
    points (\texttt{canns-gui} and a \texttt{canns} CLI
    dispatcher) and end-to-end workflows driven by configuration
    files.  The \textbf{Functional} layer hosts
    \texttt{Task}, \texttt{Trainer}, \texttt{Analyzer}, and
    \texttt{Utils} modules.  The \textbf{Core Models} layer
    encapsulates all CANN variants (basic, hierarchical,
    theta-sweep, brain-inspired).  The \textbf{Foundation} layer
    is built on \pkg{BrainPy} and \pkg{JAX} for portable
    numerics, plus a \pkg{canns-lib} Rust FFI for
    performance-critical routines.  The \textbf{Hardware} layer
    shows the supported targets: CPU, GPU, and TPU.}
  \label{fig:architecture}
\end{figure}

\autoref{fig:architecture} summaries the layer hierarchy of the
toolkit.  The layers are strictly hierarchical: every cross-layer
interaction goes through the module directly above (e.g.\
\texttt{Models} consume brain-dynamics primitives from the
Foundation layer but never call Rust code directly).

\paragraph{Application / Pipeline layer.}
User-facing surface: the GUI front end
(\texttt{canns-gui}, shipped as the optional \texttt{canns[gui]}
extra) and headless configuration-driven workflows
(\code{python -m canns.pipeline.asa}).

\paragraph{Functional layer.}
The four functional modules sit above Core Models.  \texttt{Task}
generates inputs; \texttt{Trainer} updates weights; \texttt{Analyzer}
reads outputs and produces plots, statistics, or topological
descriptors; \texttt{Utils} provides I/O, configuration
dataclasses (e.g.\ the \texttt{PlotConfig} system), and shared
helpers.

\paragraph{Core Models layer.}
This layer is where the actual neural dynamics are implemented.  All models inherit from
\code{BasicModel} or \code{BrainInspiredModel} and are implemented
on top of \pkg{BrainPy}'s \code{bp.DynamicalSystem} abstraction,
so the simulation engine is identical for the 1D CANN and the
hierarchical grid cell network.

\paragraph{Foundation layer.}
The Python stack is built on
\pkg{BrainPy}~\cite{wang2023brainpy,wang2025integrating} and
\pkg{JAX}~\cite{jax2018github}.  BrainPy provides
\code{bp.DynamicalSystem}, just-in-time (JIT)-aware \code{bm.for\_loop}, and
\code{bm.Variable} containers that hide the
functional-programming style of JAX from model authors.  The
\pkg{canns-lib} Rust backend is exposed through a thin Python
FFI for operations that would otherwise be dominated by Python
overhead (see \S3.2).

\paragraph{Hardware layer.}
The same source code runs on CPU, GPU, and TPU.  The
\texttt{canns[cuda12]}, \texttt{canns[cuda13]}, and
\texttt{canns[tpu]} extras pull in the appropriate \pkg{BrainPy}
variant for the target platform.  \pkg{canns-lib} is CPU-only at
present and used opportunistically; simulations that require
GPU/TPU acceleration stay in the JAX layer.

\subsubsection{Models Module}\label{sec:models}

The \texttt{canns.models} package contains three sub-modules:
\texttt{basic} (continuous attractor networks with predefined
connectivity), \texttt{brain\_inspired} (networks with learnable
plasticity), and \texttt{hybrid} (placeholder for future
combinations of CANNs with deep learning architectures).

\paragraph{Basic CANN models.}
The basic CANN family implements the Wu--Amari--Wong (WAW)
model~\cite{wu2008dynamics,fung2010moving,wu2016continuous},
and the analytical workhorse of the toolkit includes:

\begin{itemize}
  \item \code{CANN1D}: a 1D ring attractor with 512 neurons on the
        circle $[0, 2\pi)$, Gaussian recurrent connectivity, and
        standard leaky membrane dynamics (see the equations
        collected below).  The canonical substrate for
        head-direction encoding~\cite{taube1990head}.
  \item \code{CANN1D\_SFA}: \code{CANN1D} augmented with a slow
        activity-dependent adaptation current; produces
        anticipative tracking dynamics as characterised
        in~\cite{mi2014spike,li2025dynamics}.
  \item \code{CANN2D}: a 2D torus attractor with factorized
        Gaussian connectivity, producing bump attractors that tile
        the torus.  The canonical substrate for place
        fields~\cite{o1971hippocampus}.
  \item \code{CANN2D\_SFA}: \code{CANN2D} with SFA, enabling 2D
        traveling waves useful for spatial sequence generation.
\end{itemize}

The shared membrane and connectivity equations for the basic
CANN family are
\begin{equation}
  \begin{aligned}
    W_{ij} &= \tfrac{J_0}{\sqrt{2\pi}\,a}\,
              \exp\!\bigl(-d_{ij}^2 / 2a^2\bigr), \\
    \tau\,\dot u_i &= -u_i + \sum_j W_{ij}\,r_j + I_i, \\
    r_i &= \frac{[u_i]_+^2}{1 + k\sum_j [u_j]_+^2},
  \end{aligned}
  \quad\text{where }[u]_+ \equiv \max(u,0).
  \label{eq:cann-dynamics}
\end{equation}
where $W_{ij}$ is the recurrent weight between neurons $i$ and
$j$, $d_{ij}$ is the feature-space distance, $u_i$ is the
membrane potential, $r_i$ the firing rate, and $I_i$ the
external input.  The squaring rectification in the numerator,
combined with the divisive normalisation in the denominator, is
the canonical form of the Wu--Amari--Wong model
\citep{amari1977dynamics,wu2008dynamics,wu2016continuous}: it
makes steady-state tuning curves contrast-invariant, i.e.\ scaled
uniformly by the input strength.  In the \code{CANN1D\_SFA} and
\code{CANN2D\_SFA} variants a slow adaptation current $a_i$ is
added on top of Eq.~\eqref{eq:cann-dynamics}: it enters the
membrane equation as an additive $-a_i$ term and is itself
driven by the membrane potential via \eqref{eq:sfa}, producing
the anticipative tracking dynamics analysed in
\citet{mi2014spike,li2025dynamics} and made explicit in
\eqref{eq:sfa}.  Equation~\eqref{eq:cann-dynamics} is the
canonical reference for the basic CANN dynamics.

All basic models inherit from \code{BasicModel}, which provides
the connection-matrix builder (\texttt{make\_conn}), the
position-to-input encoder
(\texttt{get\_stimulus\_by\_pos}), the state initializer
(\texttt{init\_state}), and the diagnostic properties
(\code{self.x}, \code{self.rho}) that analyzers rely on.  Any
1D ring model with the same signature can be swapped into a
\texttt{SmoothTracking1D} task without further changes.

\paragraph{Hierarchical path integration models.}
The hierarchical path-integration family~\cite{chu2025localized}
combines a recurrent grid-cell attractor with a feedforward
place-cell read-out: \code{GaussRecUnits} (recurrent unit with
Gaussian connectivity that anchors each spatial module),
\code{BandCell} (1D path-integration unit used as a building
block for low-resolution modules), \code{GridCell} (single
multi-scale grid cell
module~\cite{hafting2005microstructure,burak2009accurate} that
integrates velocity inputs along a 2D lattice), and
\code{HierarchicalPathIntegrationModel} (full hierarchy with a
base recurrent grid-cell module that path-integrates self-motion
signals, plus \code{HierarchicalNetwork} that stacks multiple
modules~\cite{mcnaughton2006path,samsonovich1997path} at
different spatial scales together with a place-cell read-out).
These classes expose a shared \code{heal\_network()} method that
heals the connectivity after a parameter change---critical for
stable long-horizon path integration.

\paragraph{Theta-sweep models.}
Theta-sweep models reproduce the alternating forward/reverse
sweeps of place fields within individual theta
cycles~\cite{wang2015theta}.  They are built on top of CANN
primitives with a SFA-like adaptation current:
\code{DirectionCellNetwork} (head-direction implementation; with
SFA, the bump oscillates around the external heading and
produces phase precession signatures consistent
with~\cite{ji2025phase,ji2025systems}), \code{GridCellNetwork} (2D
grid modules reproducing the alternating sweep of multiple grid
fields within a theta cycle), and \code{PlaceCellNetwork} (reads
out from a grid-cell network and produces place-field sweeps of
the kind reported in~\cite{chu2024firing}).  All three classes
share the same underlying CANN1D / CANN2D infrastructure and
differ only in input encoding and readout schemes---a natural
test-bed for studying how a single circuit mechanism generalises
across cell types.

\paragraph{Brain-inspired models.}
The brain-inspired sub-module hosts networks whose connection
weights evolve under local plasticity rules.  All classes
inherit from \code{BrainInspiredModel} and expose a
\code{weight\_attr} property so the \texttt{Trainer} framework
can locate and update the weight tensor:
\code{AmariHopfieldNetwork} (the Amari--Hopfield associative
memory
network~\cite{amari1977neural,hopfield1982neural}, storing $P$
binary patterns via a generalized Hebbian
rule~\cite{hebb2005organization} and retrieving them via
asynchronous sign updates; supports continuous and discrete
dynamics), \code{LinearLayer} (one-layer linear readout with a
learnable weight matrix; the substrate for Oja's
rule~\cite{oja1982simplified}, Sanger's
rule~\cite{sanger1989optimal}, BCM~\cite{bienenstock1982theory},
and the basic Hebbian trainer), and \code{SpikingLayer} (layer
of leaky integrate-and-fire neurons with optional STDP synapses;
the bridge to spiking-network research).

\subsubsection{Task Module}

The \texttt{canns.task} module generates experimental paradigms
and input data streams.  Tasks fall into two broad categories.

\paragraph{Tracking tasks} drive a network with an external
stimulus that the network's bump must follow.  They are the
canonical test bed for CANN dynamics: a static stimulus probes
attractor stability, a smoothly moving stimulus probes tracking
bandwidth, and a brief noisy stimulus probes pattern completion.
The toolkit ships \code{PopulationCoding} (static input),
\code{TemplateMatching} (brief noisy probe), and the most common
\code{SmoothTracking1D} (continuous tracking on a ring); a 2D
variant is under development.  Tracking tasks are coupled to a
model instance at construction time
(\code{task = SmoothTracking1D(cann\_instance=cann,\dots)}) so
the task can call \code{model.get\_stimulus\_by\_pos()} to convert
feature coordinates into neural-space inputs.

\paragraph{Navigation tasks} provide the network with self-motion
signals (velocity, heading) rather than direct position, in
closed-loop and open-loop variants
(\texttt{canns.task.closed\_loop\_navigation},
\texttt{canns.task.open\_loop\_navigation}).  Navigation tasks do
not require a model instance; they expose \texttt{velocity},
\texttt{position}, and other trajectory fields and let the user
decide how to convert them into neural inputs.  This decoupling
is deliberate: the same trajectory can drive a velocity-based
grid cell model~\cite{burak2009accurate}, a place cell
readout~\cite{o1971hippocampus}, or a custom head-direction
decoder.

All tasks expose a uniform \code{get\_data()} method, a
\code{save}/\code{load} pair for reproducible serialisation, and
read the global time step via \code{bm.get\_dt()}.

\subsubsection{Analyzer Module}

The \texttt{canns.analyzer} package is the largest module by line
count.  It groups six sub-modules by function: model metrics
(\texttt{metrics.*}), visualization (\texttt{visualization.*}),
experimental / synthetic data analysis (\texttt{data.*},
including cell classification and the full ASA pipeline),
slow-point / fixed-point analysis (\texttt{slow\_points.*}),
model-specific tools (\texttt{model\_specific.*}), and
higher-level composed workflows (\texttt{workflows.*}).  The
self-description in \texttt{canns.analyzer.\_\_init\_\_} is
``analysis helpers such as model metrics, visualization,
experimental or synthetic data analysis, slow-point
(fixed-point) analysis, and model-specific tools''.

\paragraph{Model Metrics.}
Computational analysis with no Matplotlib dependency.
\texttt{metrics.spatial\_metrics} computes grid score, spatial
autocorrelation, grid spacing, and Gaussian-smoothed firing
fields; \texttt{metrics.systematic\_ratemap} samples the 2D
position grid in a Burak--Fiete-style structured sweep so that
rate maps have 100\% spatial coverage and preserve stable CANN
dynamics~\citep{burak2009accurate}; \texttt{metrics.utils}
provides spike-train conversion helpers.  This sub-module is
the workhorse of any pipeline that needs numerical descriptors
of neural activity without rendering a figure.

\paragraph{Visualization.}
Visualises simulation outputs.  It is \emph{array-in, figure-out}:
the functions accept firing rates or membrane potentials as
\texttt{(time, neurons)} arrays and produce Matplotlib figures or
animations.  The model itself is \emph{not} required at plot time,
which makes the analyzer trivially compatible with any model
that produces arrays of the expected shape.  Core capabilities
include activity visualisation (\code{animate\_dynamics},
\code{plot\_network\_state}, \code{plot\_bump\_trajectory}),
energy landscape (\code{energy\_landscape\_1d} and
\code{energy\_landscape\_2d} show the
$E(\mathbf{r}) = -\frac{1}{2}\,\mathbf{r}^\top W \mathbf{r}
+ \mathbf{I}^\top \mathbf{r}$ surface, where $\mathbf{I}$ is the
external input; this is the cleanest way to inspect
attractor basin structure), connectivity
(\code{plot\_weight\_matrix}, \code{plot\_connection\_profile},
which show the Mexican-hat recurrent kernel for debugging
\code{make\_conn()}), raster / firing-rate plots, and tuning
curves.  A \code{PlotConfig} dataclass centralises figure size,
color map, and animation speed and is passed across multiple
plots to keep a visualisation set coherent.

\paragraph{Data Analysis.}
Operates on experimental recordings (spike trains, firing-rate
estimates) and on virtual data, and is the home of the full ASA
pipeline machinery.  \texttt{data.cell\_classification} provides
cell-classification utilities (GridScore, head-direction tuning
via mean vector length and Rayleigh test, grid-module clustering
via Leiden community detection).  \texttt{data.asa} provides
the ASA pipeline components used in \S4: spike-train embedding
(\code{embed\_spike\_trains}), time- and spatial-TDA
(\code{tda\_vis}, \code{spatial\_tda})
\citep{carlsson2009topology,edelsbrunner2010computational},
circular-coordinate decoding
(\code{decode\_circular\_coordinates\_multi}), CohoMap /
CohoSpace / path comparison
(\code{cohomap}, \code{cohospace},
\code{plot\_path\_compare\_1d/2d}), firing-rate heatmaps
(\code{fr}), and population activity analysis (bump position
from a population vector, Gaussian fits to firing fields,
decoded position over time).  The \pkg{canns-lib} Ripser module
(see \S3.2) plugs into the inner loop; the semantic interface
(Betti numbers, barcode, persistence diagrams) is identical to
the pure-Python reference, so all downstream consumers
(CohoMap, CohoSpace, PathCompare, CohoScore in the ASA
Pipeline) work transparently with either backend.

\paragraph{Slow-Point / Fixed-Point Analysis.}
Applies techniques from dynamical-systems analysis of recurrent
networks~\cite{sussillo2013opening} to CANNs.  Exposes a
\code{slow\_points.finder} that solves
$\dot{\mathbf{u}} = f(\mathbf{u}) = 0$ for fixed points using
JAX-backed numerical optimisation similar in spirit to
FixedPointFinder~\cite{golub2018fixedpointfinder}, classifies
them by Jacobian eigenvalues (stable / unstable / saddle), and
produces diagnostic plots (\code{plot\_fixed\_points\_2d/3d});
\texttt{slow\_points.checkpoint} provides save / load for
restarting long searches.

\paragraph{Model-Specific Tools.}
Holds analyzers that only make sense for one model class.
Currently \texttt{model\_specific.hopfield} provides analysis
helpers for the Amari--Hopfield associative memory
(pattern-overlap dynamics, energy diagnostics), the natural
complement to \code{canns.models.brain\_inspired.AmariHopfieldNetwork}.

\paragraph{Higher-Level Workflows.}
Composed analysis workflows that wire lower-level analyzer
modules together.  \texttt{workflows.auto\_grid\_threshold}
sweeps grid-score-ranked neuron subsets to identify a clean
population for downstream TDA or coho analysis, with summaries
of shuffle / torus topology at each candidate threshold;
\texttt{workflows.phase\_center\_comparison} compares phase
centers across sessions or conditions under torus-aware
displacement, returning paired black/red phase-center plots and
displacement summaries.

\subsubsection{Trainer Module}

The \texttt{canns.trainer} module implements the local,
activity-dependent plasticity rules that drive the
brain-inspired models.  Five trainers ship in the box:
\code{HebbianTrainer} (outer-product
rule~\cite{hebb2005organization}), \code{OjaTrainer}
(normalised Hebbian rule for principal-component
extraction~\cite{oja1982simplified}), \code{SangerTrainer}
(generalised Hebbian rule for multiple principal
components~\cite{sanger1989optimal}), \code{BCMTrainer}
(Bienenstock--Cooper--Munro
rule~\cite{bienenstock1982theory,bi1998synaptic}), and
\code{STDPTrainer} (spike-timing-dependent plasticity for
\code{SpikingLayer}).  Each trainer pairs with a specific model
class through the \code{weight\_attr} contract and is
agnostic to the underlying network topology.  All trainers
support both \texttt{fit\_online} (per-step update) and
\texttt{fit\_offline} (batch) regimes; the same \code{Trainer}
can be reused across models with no changes.

\subsection{Extensibility}

Extending \pkg{canns} is a matter of subclassing and overriding
a few methods, because every component plugs into a small number
of base-class contracts.  To add a new basic CANN model,
subclass \code{BasicModel} (or \code{BasicModelGroup} for
multi-population models), call the parent constructor with the
total neuron count, and implement \code{make\_conn()},
\code{get\_stimulus\_by\_pos()}, \code{init\_state()}, and
\code{update()}.  The \texttt{CANN1D} implementation in
\texttt{src/canns/models/basic/cann.py} is the smallest
end-to-end reference.  To add a new brain-inspired model or
trainer, subclass \code{BrainInspiredModel}, register state and
weight variables with \code{bm.Variable}, implement
\code{update()}, and expose a \code{weight\_attr} property if
needed; pair the model with a \code{Trainer} subclass that
implements \code{train()} and \code{predict()}; the
\texttt{AmariHopfieldNetwork}, \texttt{OjaTrainer}, and
\texttt{HebbianTrainer} classes are the canonical short
references.  Tasks, analyzer functions, and pipelines are added
through their respective base classes; see the developer
documentation for the full extension contract.

\subsection{\pkg{canns-lib} (Rust): the Acceleration Layer}\label{sec:canns-lib}

\pkg{canns-lib} is a companion Rust crate that exposes a
performance-critical subset of the CANN routines to Python
through a thin PyO3 FFI layer.  The Python module API is
intentionally a superset-compatible drop-in: any code that works
with the pure-Python implementation works with the Rust
implementation when the wheel is installed, with no conditional
logic at the calling site.  The motivation is that several
bottleneck operations---persistent homology, long spatial
navigation trajectories, bulk task generation---have low
arithmetic intensity per Python object and are dominated by
interpreter overhead.  The key performance numbers, as reported
in the \pkg{canns-lib} documentation and re-measured on the
present hardware, are summarised in
\autoref{tab:canns-lib-perf}.

\begin{table}[ht]
  \centering
  \caption{Performance of \pkg{canns-lib} on representative
    workloads.  Speedup is the geometric mean over the benchmark
    suite (Ripser: 54 datasets, spatial: $10^2$--$10^6$ step
    sweeps), with the peak ratio in parentheses for Ripser.
    Both benchmarks are CPU-only on a single thread
    (Apple M2 Pro, 16\,GB RAM).  ``ripser.py'' is the
    Cython/\texttt{C++} reference implementation; ``pure-Python
    RatInABox'' is the original \code{RatInABox} environment.}
  \label{tab:canns-lib-perf}
  \begin{tabular}{l c c c}
    \hline
    \textbf{Module} & \textbf{Baseline} &
    \textbf{\pkg{canns-lib}} & \textbf{Speedup} \\
    \hline
    Ripser
      & ripser.py (Cython/\texttt{C++}) & Rust port
      & 1.13$\times$ avg (1.82$\times$ peak) \\
    Spatial navigation
      & pure-Python RatInABox & Rust + PyO3 env
      & $\sim$700$\times$ \\
    \hline
  \end{tabular}
\end{table}

\paragraph{Ripser acceleration.}
The Rust Ripser module is a drop-in replacement for
\texttt{ripser.py} (the Cython wrapper around the C++
reference Ripser implementation).  It supports coefficient rings,
homology in dimensions 0, 1, and 2, reduction-order statistics,
multiple distance metrics (Euclidean, Manhattan, cosine, custom),
sparse distance matrices, and optional representative-cocycle
output.  The internal algorithmic improvements over
\texttt{ripser.py} are row-by-row edge generation, binary
search on sparse columns, a structure-of-arrays memory layout,
and Rayon-based multi-threading.  The 1.13$\times$ geometric-mean
speedup reflects the fact that the comparison baseline
(\texttt{ripser.py}) is already a tight C++ implementation; the
Rust version wins by eliminating Python interpreter overhead
in the filtration bookkeeping and by parallelising the dense
column reductions.  Memory usage is essentially identical
(1.01$\times$ ratio).  All speedup numbers are from the
internal benchmark suite in the \pkg{canns-lib} repository
(54 datasets spanning random Gaussian, two-moons, circle,
torus, and synthetic grid-cell manifolds, dimensions 2--6,
point-cloud sizes $200$--$10{,}000$).

\paragraph{Spatial navigation acceleration.}
Long-horizon path integration simulations spend most of their
wall time in the environment simulator (collision checking,
boundary handling, sensory sampling) rather than in the network
update itself.  The Rust port of the \code{RatInABox}-style
spatial navigation module delivers a $\sim$700$\times$ speedup
on long trajectories (10--30 minutes of simulated time at
1\,kHz update rate)---the difference between an overnight run
and a coffee break.  The implementation is a
\pkg{PyO3}-accelerated re-implementation of \code{RatInABox}
that retains the same \texttt{Environment} / \texttt{Agent} API,
so existing Python code that uses \code{RatInABox} can switch
by changing only the import statement.  The accelerated
implementation supports solid and periodic boundaries, arbitrary
polygons, holes, and thigmotaxis wall-following; the Rust core
is exposed to Python through the \texttt{canns\_lib.spatial}
sub-package.  At $10^6$ steps the \pkg{canns-lib} runtime is
0.27\,s versus 192.8\,s for the pure-Python \code{RatInABox}
reference (a 726$\times$ speedup on this configuration).

\paragraph{Module coverage and roadmap.}
The two \pkg{canns-lib} modules shipped today are
\texttt{canns\_lib.ripser} and \texttt{canns\_lib.spatial}; both
are released and versioned on PyPI (\texttt{canns-lib\,$\geq$0.6.2}),
both are declared hard dependencies of \pkg{canns}, and both are
in the default install.  Approximate nearest neighbours, an
accelerated CANN dynamics-computation kernel, and a
shared-memory parallel batch evaluator are planned for future
releases; the \pkg{canns-lib} repository tracks these in its
issue tracker.  When the wheel is unavailable the toolkit
gracefully falls back to a pure-Python reference for the Ripser
path; the spatial path is not currently available without the
wheel.

\paragraph{Integration story.}
\pkg{canns-lib} is released as a separate PyPI package and lives
in its own sister repository
(\url{https://github.com/Routhleck/canns-lib}).  When the wheel
is available, Python's import machinery picks up the Rust
implementation; the user-visible behaviour is identical to
the pure-Python fallback for the Ripser path.  The spatial
path requires the wheel.

\begin{figure}[ht]
  \centering
  \begin{subfigure}[t]{0.85\textwidth}
    \includegraphics[width=\textwidth]{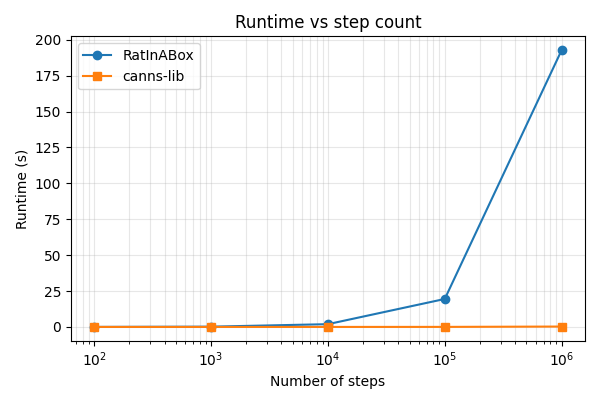}
    \caption{Runtime vs.\ number of integration steps
      ($10^2$--$10^6$).  Log-log axes compare pure-Python
      \texttt{RatInABox} (orange) with the \pkg{canns-lib} Rust
      port (blue).  At $10^6$ steps the Rust port finishes in
      $\approx 0.27$\,s versus $192.8$\,s for the pure-Python
      reference.}
    \label{fig:canns-lib-spatial-runtime}
  \end{subfigure}
  \vspace{0.5em}
  \\
  \begin{subfigure}[t]{0.85\textwidth}
    \includegraphics[width=\textwidth]{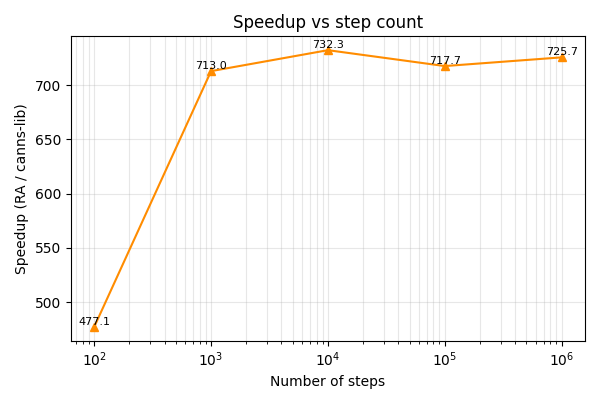}
    \caption{Speedup ratio (RatInABox\,/\,canns-lib) over the
      same step-count sweep.  The geometric mean across
      $10^2$--$10^6$ steps is $\approx 700\times$ (individual
      data-point labels reproduced from the benchmark script).}
    \label{fig:canns-lib-spatial-speedup}
  \end{subfigure}
  \caption{\pkg{canns-lib} spatial-navigation benchmark.  Panel
    (a) shows raw runtimes and panel (b) shows the per-step-count
    speedup.  Both plots are produced by
    \texttt{canns-lib/benchmarks/spatial/step\_scaling\_benchmark.py}
    with default parameters (\texttt{dt=0.02}, unit-square
    environment with four solid walls, agent start $[0.5,0.5]$).
    Methodology, hardware, and per-scenario numbers are
    documented in the \pkg{canns-lib} README.}
  \label{fig:canns-lib-spatial}
\end{figure}

\begin{figure}[ht]
  \centering
  \includegraphics[width=0.85\textwidth]{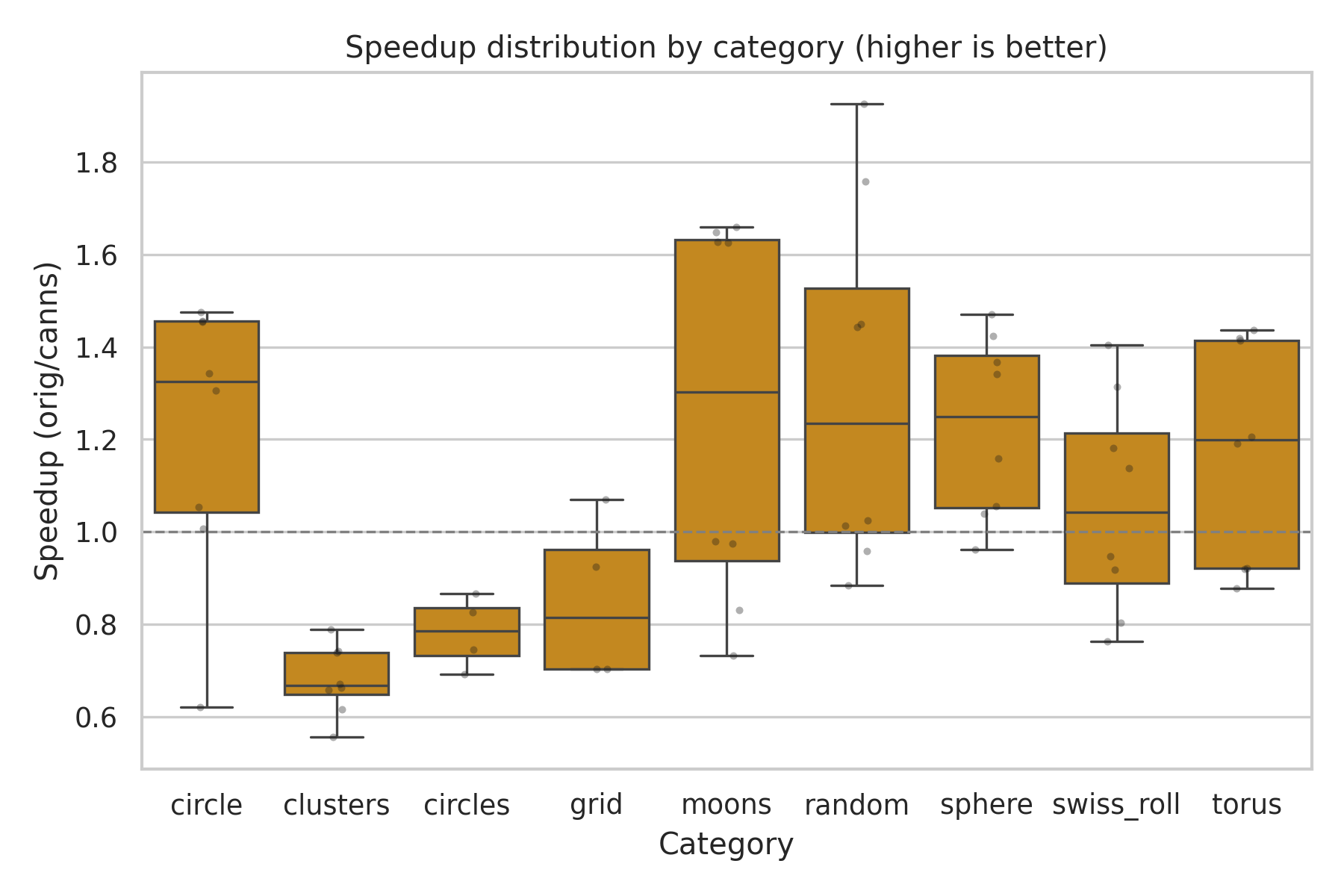}
  \caption{Speedup of the \pkg{canns-lib} Rust Ripser port
    versus the \texttt{ripser.py} Cython baseline, broken down by
    point-cloud category.  Across the 54-dataset benchmark
    suite (random Gaussian, two-moons, circle, torus, grid,
    clusters, swiss roll, concentric circles) the geometric-mean
    speedup is $1.13\times$ and the peak is $1.82\times$.  Box
    plots show the distribution within each category; the dashed
    line at 1.0 marks parity.  Accuracy is bit-exact with the
    baseline: bottleneck distance between matched persistence
    pairs stays well below the $0.02$ tolerance and the H$_0$/H$_1$
    match rate is 100\% across all 54 datasets, so the Rust port
    is a drop-in replacement.  Produced by
    \texttt{canns-lib/benchmarks/ripser/comprehensive\_benchmark.py};
    per-run JSON and logs ship with the \pkg{canns-lib}
    repository.}
  \label{fig:canns-lib-ripser}
\end{figure}

\paragraph{Benchmark figures.}
\autoref{fig:canns-lib-spatial} and \autoref{fig:canns-lib-ripser}
plot the raw benchmark output produced by the scripts in
\pkg{canns-lib}'s \texttt{benchmarks/} directory (see the
\texttt{canns-lib} repository for the exact methodology and
CLI flags).

\subsection{ASA: the Experimental-Data Layer}

The third component is the Attractor Structure Analyzer (ASA), a
pipeline that takes the toolkit's analysis layer one step
further: given a pre-recorded experimental dataset
(spike times or firing-rate matrices from a population of
$N$ simultaneously recorded cells), ASA turns the question
\emph{``does this population code carry a ring, torus, or
sphere signature?''} into a reproducible computation.  The
output is a set of topological descriptors (Betti numbers,
persistent barcode, circular coordinates) plus downstream
visualisations (CohoMap, CohoSpace, PathCompare) that map the
abstract descriptors back onto the cell population and the
animal's behaviour.

ASA is a thin \emph{domain application} built on top of the
\pkg{canns} library: it consumes the same Data Analyzer
utilities (cell classification, tuning-curve analysis), the
same Rust-accelerated Ripser module, and the same plotting
back end.  The added value of ASA is not new analysis primitives
(those come from the TDA
literature~\cite{carlsson2009topology,edelsbrunner2010computational,Vaupel2023Duality})
but a single coherent pipeline that wires them together with
sensible defaults, a graphical front end, and a result
directory that bundles the JSON summary, the figure files, and
the configuration that produced them.  All of the
methodological detail (preprocessing, persistent homology,
persistent cohomology, CohoMap/CohoSpace/PathCompare, CohoScore,
auto-grid threshold, GUI implementation, and result-management
file layout) is in \S4; the present subsection only tells the
reader where ASA fits in the ecosystem and how to launch it.

\paragraph{Launching ASA.}
The canonical entry point is the graphical front end
(\texttt{canns-gui}, shipped as the \texttt{canns[gui]} extra),
which exposes preprocessing, analysis, and visualisation in a
single \pkg{PySide6} window.  Headless invocation is supported
via \code{python -m canns.pipeline.asa --config config.yaml}
for batch experiments.  Both routes produce a
\texttt{<workdir>/Results/<dataset\_id>/} directory
containing a JSON summary, the analysis figures, and the
configuration used.

\subsection{Reproducibility}

Every published result in \S5 and \S6 of this paper can be
regenerated from the public repository with a single command.
The \pkg{canns} source code, the \pkg{canns-lib} Rust backend,
and the ASA pipeline are released as open-source software; the
source repositories track the current release, and the public
Docker images pin a reproducible Python environment.  Example notebooks that reproduce every
figure in \S5 and \S6 ship under \texttt{examples/} in the
\pkg{canns} repository, and a single \texttt{make reproduce}
command in the repository root re-runs the full suite.

%% file: sections/asa.tex

\section{The ASA Pipeline: Engineering Attractor-Structure Analysis}
\label{sec:asa}

\subsection{Motivation and Scope}
\label{sec:asa:motivation}

CANN theory predicts that the collective activity of a neural
population encoding a continuous variable is confined to a
low-dimensional attractor manifold.  ASA is not restricted to the
two-dimensional torus.  A one-dimensional periodic code, such as head
direction, a band-cell phase, or a 1D CANN, is expected to form a
ring $S^1$.  A two-dimensional periodic code, such as a grid-cell
module or a 2D CANN, is expected to form a torus
$T^2=S^1\times S^1$.  In recordings, however, the manifold is never
observed directly.  The experimenter observes spike times or rate
estimates from $N$ neurons, behavioural variables such as position
and speed, and a large number of nuisance factors: non-uniform
sampling, heterogeneous firing rates, inactive units, speed-dependent
modulation, and partial spatial coverage.

The Attractor Structure Analyzer (ASA) is the component of the
\pkg{canns} ecosystem that turns these observations into reproducible
attractor-structure analyses.  ASA is not a single persistent-homology
script.  It is a modular engineering system that combines a unified
\texttt{.npz} data format, spike/rate preprocessing, time-indexed and
spatially indexed point-cloud construction, and persistent homology,
shuffle controls, persistent cohomology decoding, CohoMap/EcohoMap,
CohoSpace/EcohoSpace, PathCompare, GridScore, CohoScore,
module-level workflows, a graphical interface, a terminal interface,
and cache-aware result management.

The core mathematical object in the time-indexed route is the
population activity vector
\begin{equation}
  r(t)=\bigl[r_1(t),r_2(t),\ldots,r_N(t)\bigr]^\top\in\mathbb{R}^N ,
  \label{eq:asa-pop-vector}
\end{equation}
sampled at many time points $t$.  The corresponding point cloud is
\begin{equation}
  X_t=\{r(t_j):j=1,\ldots,T\}\subset\mathbb{R}^N ,
  \label{eq:asa-time-cloud}
\end{equation}
which asks whether the neural trajectory visits a low-dimensional
manifold.  ASA also supports a spatial route in which the data are
first aggregated by physical position $x$ and the point cloud becomes
\begin{equation}
  X_x=\{r(x_\ell):\ell=1,\ldots,L\}\subset\mathbb{R}^N .
  \label{eq:asa-spatial-cloud}
\end{equation}
This route asks a different question: whether the map from physical
space to population representation is topologically structured.
Keeping $r(t)$ and $r(x)$ separate is essential, because temporal
continuity and spatial representational geometry are related but not
identical scientific claims.  In either route, ASA can decode one
circular coordinate for a ring-like attractor or two circular
coordinates for a toroidal attractor.

\subsection{Unified Data Model and Preprocessing}
\label{sec:asa:input}

ASA uses a single NumPy \texttt{.npz} container as its interchange
format.  The required key is \texttt{spike}; optional but commonly
used keys include \texttt{t}, \texttt{x}, \texttt{y},
\texttt{trial\_id}, and \texttt{step\_id}.  The \texttt{spike}
field can take two forms.  In experimental spike-time data it is a
per-neuron spike-train structure, which ASA converts to a continuous
rate matrix by binning and Gaussian smoothing.  In simulated or
preprocessed datasets it may already be a dense matrix
$R\in\mathbb{R}^{T\times N}$; in this case ASA follows the
\emph{skip-embed} path and uses the matrix directly.

For spike-time input, the embedding step approximates
\begin{equation}
  r_i(t)=\int K_\sigma(t-u)\sum_j\delta(u-t_i^j)\,du ,
  \label{eq:asa-spike-embedding}
\end{equation}
where $K_\sigma$ is a Gaussian kernel and $t_i^j$ are the spike
times of neuron $i$.  The kernel width $\sigma$ (in time units) is
chosen by the user; a typical value for neural spike data is
$\sigma\approx25\,$ms.  The resulting rate matrix is aligned with
behavioural traces, optionally speed-filtered, standardized across
neurons, and projected to a low-dimensional principal component analysis
(PCA) space (typically 6 dimensions for neural population data) before TDA.
The preprocessing parameters, input hash, and output artefacts are
saved together, so a figure can be traced back to its data and
configuration.

ASA exposes three analysis routes from the same container:
\begin{enumerate}
  \item \emph{Time-indexed trajectory analysis}, using $X_t$ for
    persistent homology, circular-coordinate decoding, CohoMap,
    CohoSpace, and PathCompare.
  \item \emph{Spatial representation analysis}, using $X_x$ built
    from firing-rate maps or spatial bins, corresponding to the
    Fig.~4C-style population-representation analysis used in recent
    grid-cell topology work.
  \item \emph{Neuron/module feature analysis}, using single-cell
    firing-rate maps, GridScore, CohoScore, phase centers, and
    module labels to select and compare cell subsets.
\end{enumerate}

\subsection{Persistent Homology}
\label{sec:asa:homology}

Given a point cloud $X=\{x_i\}_{i=1}^M$ and a distance metric
$d(\cdot,\cdot)$, ASA computes persistent homology through the
Vietoris--Rips filtration.  At scale $\epsilon$, the complex
\begin{equation}
  \mathrm{VR}_\epsilon(X)
  =
  \left\{
    \sigma\subseteq X:
    d(x_a,x_b)\leq \epsilon
    \;\; \textrm{for all } x_a,x_b\in\sigma
  \right\}
  \label{eq:asa-vr}
\end{equation}
contains all simplices whose vertices are pairwise closer than
$\epsilon$.  As $\epsilon$ increases, the complexes form a nested
filtration.  Homology tracks topological features across this
filtration.  The $k$-th homology group is
\begin{equation}
  H_k(K;F)=\ker\partial_k/\operatorname{im}\partial_{k+1},
  \label{eq:asa-homology}
\end{equation}
and its rank $\beta_k$ is the $k$-th Betti number.  In the present
setting, $\beta_0$ counts connected components, $\beta_1$ counts
independent loops, and $\beta_2$ counts enclosed two-dimensional
cavities.  A ring-like attractor has the ideal signature
\begin{equation}
  S^1:\qquad (\beta_0,\beta_1,\beta_2)=(1,1,0),
  \label{eq:asa-110}
\end{equation}
whereas a toroidal attractor has the ideal signature
\begin{equation}
  T^2:\qquad (\beta_0,\beta_1,\beta_2)=(1,2,1).
  \label{eq:asa-121}
\end{equation}
A contractible cloud has $(1,0,0)$.  Thus a single stable $H_1$
bar can be the correct target for HD, band-cell, or 1D CANN
analyses, while two stable $H_1$ bars plus a compatible $H_2$
feature are expected only when the scientific hypothesis is toroidal.

The output is represented as persistence intervals $(b,d)$, where
$b$ is the birth scale and $d$ is the death scale.  ASA visualizes
these intervals as barcodes and persistence diagrams.  Long bars are
candidate attractor signatures, but they are not interpreted in
isolation: sampling density, finite trajectory coverage, PCA
truncation, and correlated firing can also generate long-lived
features.  ASA therefore couples the barcode with explicit null
models.

The persistent homology computation uses the Vietoris--Rips filtration
via the \pkg{Ripser} algorithm (as implemented in \pkg{canns-lib}) with
coefficients over $\mathbb{Z}/2\mathbb{Z}$.  The maximum homology
dimension \texttt{maxdim} is specified by the user: \texttt{maxdim=1}
is sufficient for ring-like attractor analyses, while
\texttt{maxdim=2} is required to detect the $H_2$ cavity that
distinguishes a torus from two independent rings.  The filtration scale
range is automatically determined from the pairwise distance distribution
of the point cloud; the default behaviour scales the maximum filtration
value to twice the median pairwise distance.

\subsection{Shuffle Null Models}
\label{sec:asa:shuffle}

For time-indexed analyses ASA uses circular-shift shuffle controls.
Each neuron is shifted independently along time, preserving its
autocorrelation and marginal firing-rate distribution while
destroying precise inter-neuron co-activation.  The complete
preprocessing and TDA chain is then rerun on each shuffled dataset.
The resulting distribution of maximum lifetimes provides an
empirical null envelope for the observed barcode.  In figures, ASA
can draw this null envelope as a grey shadow behind the real green
barcode, making the comparison visually direct.

For spatial representation analyses ASA uses a different null.  The
valid occupied spatial bins are held fixed, and each neuron's
spatial rate map is permuted within that valid mask.  This preserves
the set of sampled locations and each neuron's rate distribution, but
breaks the coordinated population representation across neurons.  The
spatial null is therefore matched to $r(x)$ rather than $r(t)$.

Shuffle results should be interpreted as empirical controls, not as
absolute proofs.  With $S$ shuffles, the smallest attainable
empirical $p$ value is $1/(S+1)$, so a 50-shuffle run is useful for
screening and visualization, whereas stronger claims benefit from
more shuffles and explicit reporting of the null percentile used.
For classification into ``strong torus / partial / weak'' categories
(Sec.~\ref{sec:asa-selected20}), the 99.9th percentile ($q=0.999$)
of the shuffle lifetime distribution is used as the threshold.

\subsection{Persistent Cohomology and Circular Coordinates}
\label{sec:asa:cohomology}

Persistent homology identifies which holes are present.  Persistent
cohomology provides coordinates around those holes.  Let $C^k(K;F)$
be the $k$-cochain group of a simplicial complex $K$, with coboundary
operator $\delta^k:C^k\rightarrow C^{k+1}$.  A persistent
$H^1$ class can be represented by a cocycle on edges.  Following the
circular-coordinate construction of
\citet{desilva2011persistent}, ASA lifts this cocycle to a
real-valued phase function by solving a least-squares problem of the
form
\begin{equation}
  \min_{f,n_{ij}\in\mathbb{Z}}
  \sum_{[v_i,v_j]\in E}
  \left(
    f(v_j)-f(v_i)-\alpha_{ij}-n_{ij}
  \right)^2 ,
  \label{eq:asa-cocycle-lift}
\end{equation}
where the integer variables absorb branch-cut jumps.  Reducing the
lift $f$ modulo $1$, and then rescaling by $2\pi$, gives a circular
coordinate $\theta:X\rightarrow S^1$.

More generally, ASA decodes $m$ circular coordinates,
\begin{equation}
  \Theta_m(t)=
  \bigl(\theta_1(t),\ldots,\theta_m(t)\bigr)
  \in (S^1)^m .
  \label{eq:asa-theta}
\end{equation}
The choice of $m$ is both data-driven and hypothesis-driven:
$m=1$ is used for ring-like systems such as head-direction cells,
band cells, and 1D CANNs, while $m=2$ is used for toroidal systems
such as grid-cell modules.  The decoded coordinates are stored
together with the preprocessing basis and landmark/reference
information, so downstream analyses use the same coordinate frame
rather than recomputing an unrelated embedding.

\subsection{CohoMap, EcohoMap, CohoSpace, and PathCompare}
\label{sec:asa:cohomap}

The decoded coordinates $\Theta_m(t)$ are intrinsic coordinates on the
inferred manifold.  ASA provides several interpretation modules that
relate them back to physical space and single-cell activity.

\emph{CohoMap} and \emph{EcohoMap} map cohomology-derived phase back
to behavioural or physical coordinates.  In one-dimensional mode,
ASA maps a single decoded phase $\theta(t)$ to a behavioural variable
such as head direction, band phase, or ring position.  In
two-dimensional mode, ASA bins physical space and computes circular
means of the two decoded phases.  EcohoMap avoids phase
discontinuities by working with sine/cosine representations and can
align phase bands to make the inferred circular directions easier to
interpret.  In grid-cell data, a good two-dimensional EcohoMap shows
two spatially organized phase families whose period and orientation
are consistent with grid-like coding.

\emph{CohoSpace} and \emph{EcohoSpace} move in the opposite
direction: they ask how a single neuron's firing rate is arranged on
the decoded attractor.  In two-dimensional mode, for neuron $i$, ASA
estimates
\begin{equation}
  F_i(u,v)=
  \mathbb{E}\!\left[
    r_i(t)\mid
    \theta_1(t)\in B_u,\;
    \theta_2(t)\in B_v
  \right],
  \label{eq:asa-cohospace}
\end{equation}
where $(B_u,B_v)$ are phase bins.  In one-dimensional mode the same
idea reduces to
\begin{equation}
  F_i(u)=
  \mathbb{E}\!\left[
    r_i(t)\mid
    \theta(t)\in B_u
  \right].
  \label{eq:asa-cohospace-1d}
\end{equation}
A grid cell that has many fields in physical space may become a
compact single bump in two-dimensional cohomology space, because the
repeated physical fields are periodic images of one toroidal phase
preference.  An HD or band cell can analogously become concentrated
around one preferred ring phase.

\emph{PathCompare} compares the behavioural trajectory
$(x(t),y(t))$ or a one-dimensional behavioural variable with the
decoded phase trajectory $\Theta_m(t)$ on the same time axis.  In
one-dimensional mode this is a comparison between a circular variable
and $\theta(t)$.  In two-dimensional mode it compares physical motion
with movement on the skewed torus.  ASA includes direction alignment
and scaling utilities, because different modules can encode the same
physical displacement with different phase gains.  PathCompare is
therefore a qualitative and quantitative diagnostic for whether the
decoded attractor motion respects the behavioural trajectory.

\subsection{GridScore, CohoScore, and Module Workflows}
\label{sec:asa:cellmetrics}

ASA combines classical spatial-tuning metrics with topology-aware
metrics.  GridScore is computed from the spatial autocorrelogram of a
cell's firing-rate map.  If $r_\phi$ denotes the correlation between
the autocorrelogram and its rotation by angle $\phi$, the standard
score is
\begin{equation}
  \mathrm{GridScore}
  =
  \min(r_{60},r_{120})
  -
  \max(r_{30},r_{90},r_{150}).
  \label{eq:asa-gridscore}
\end{equation}
ASA can evaluate this score across radius windows and uses it for
cell ranking, grid-module analysis, and automatic threshold searches.

CohoScore measures how concentrated a neuron's activity is in
decoded phase space.  For a set of high-activity time points
$\mathcal{M}_i$ of neuron $i$, ASA computes circular variance for
each decoded phase,
\begin{equation}
  \mathrm{Var}_{\mathrm{circ}}(\theta)
  =
  1-\left|
    \frac{1}{|\mathcal{M}_i|}
    \sum_{t\in\mathcal{M}_i} e^{\mathrm{i}\theta(t)}
  \right|,
  \label{eq:asa-circular-var}
\end{equation}
and averages over the decoded circular coordinates:
\begin{equation}
  \mathrm{CohoScore}_i
  =
  \frac{1}{m}\sum_{k=1}^{m}
  \mathrm{Var}_{\mathrm{circ}}\!\left(
    \{\theta_k(t):t\in\mathcal{M}_i\}
  \right).
  \label{eq:asa-cohoscore-m}
\end{equation}
Lower CohoScore means stronger phase concentration.  For $m=1$ this
measures concentration on a ring; for $m=2$ it measures concentration
on a torus.  Thus high GridScore and low CohoScore are expected to
co-occur in cells that strongly participate in a grid-cell toroidal
population code, while the same metric also applies to HD or band
cells in one-dimensional analyses.

The auto-grid-threshold workflow treats cell selection as a
data-driven search rather than a fixed manual cutoff.  It sorts cells
by GridScore, runs top-$k$ candidate subsets through TDA and
cohomology decoding, evaluates H$_1$ or H$_1$/H$_2$ topology
statistics, optionally refines promising candidates with shuffle
controls, and measures the GridScore--CohoScore relationship.  This
workflow is deliberately placed above the primitive analysis modules:
it composes existing ASA operations rather than introducing a
separate definition of topology.

ASA also includes workflows for comparing cohomology phase centers
across sessions or datasets.  Matched neurons can be connected on the
skewed torus, and displacements are computed with the toroidal
minimum-image convention, avoiding artificial long jumps across the
$0/2\pi$ boundary.

\subsection{Interfaces, Caching, and Provenance}
\label{sec:asa:gui}

\begin{figure}[!htbp]
  \centering
  \begin{subfigure}[t]{0.32\textwidth}
    \includegraphics[width=\textwidth]{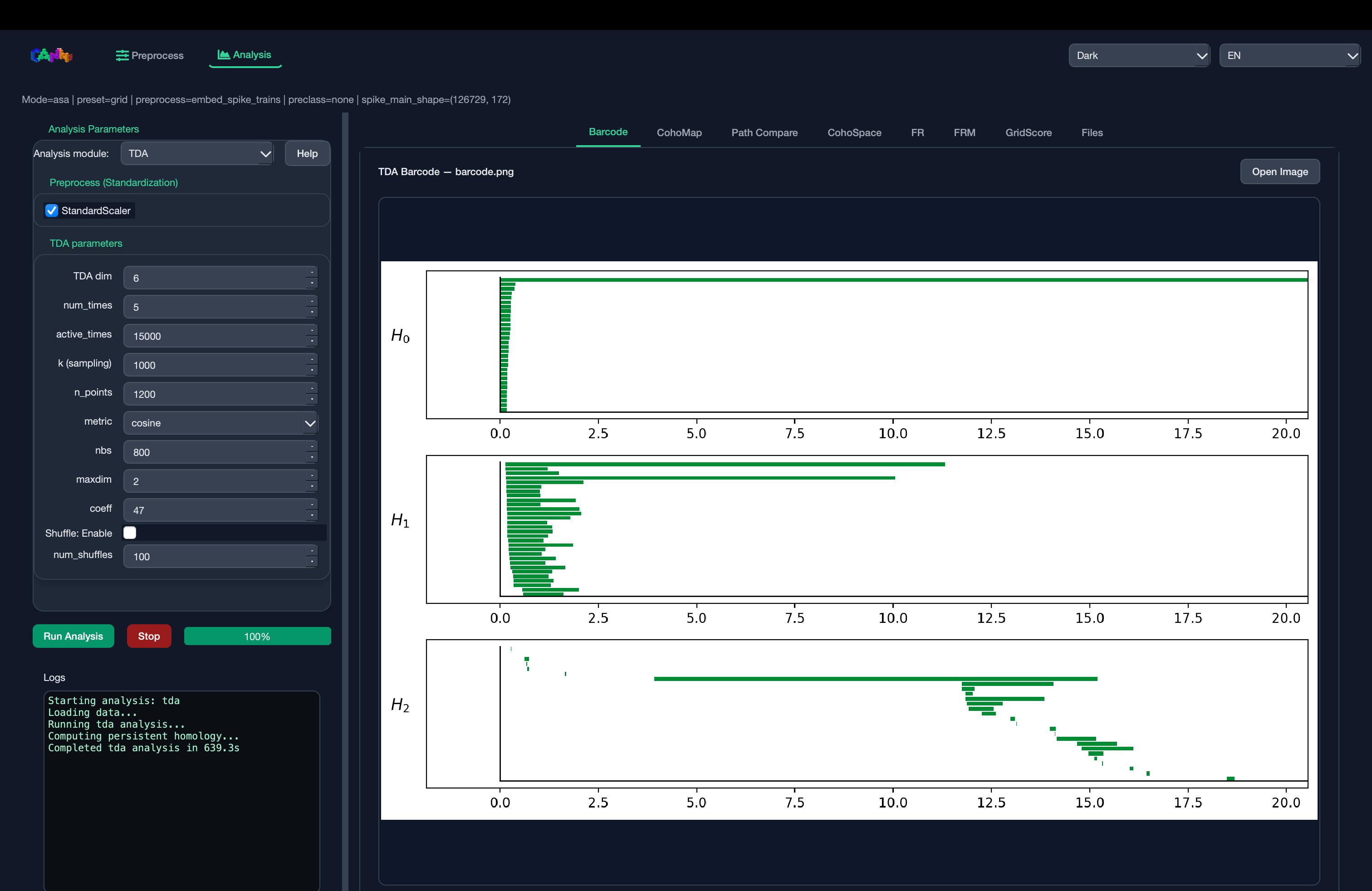}
    \caption{TDA barcode mode.}
  \end{subfigure}
  \hfill
  \begin{subfigure}[t]{0.32\textwidth}
    \includegraphics[width=\textwidth]{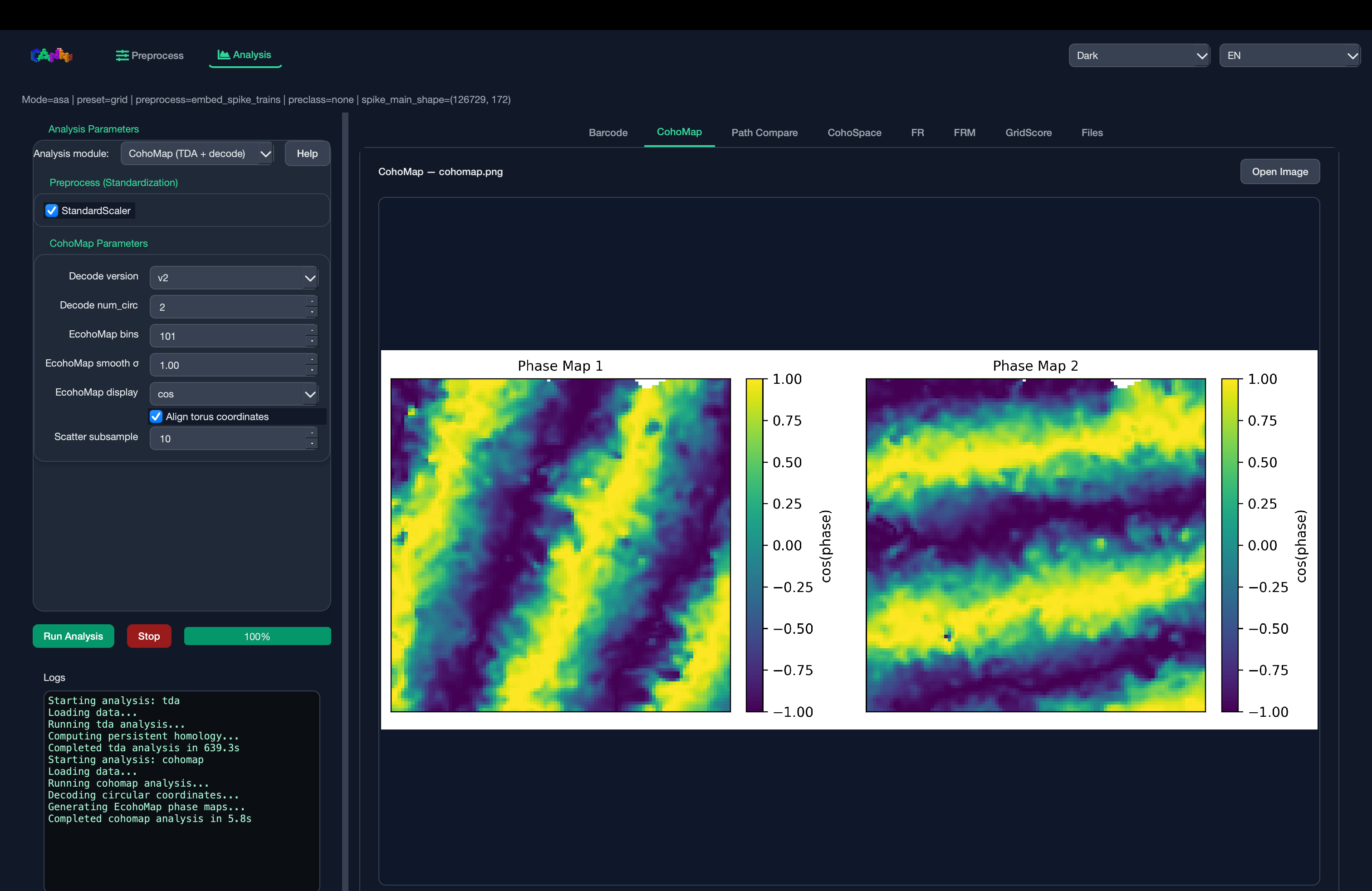}
    \caption{EcohoMap mode.}
  \end{subfigure}
  \hfill
  \begin{subfigure}[t]{0.32\textwidth}
    \includegraphics[width=\textwidth]{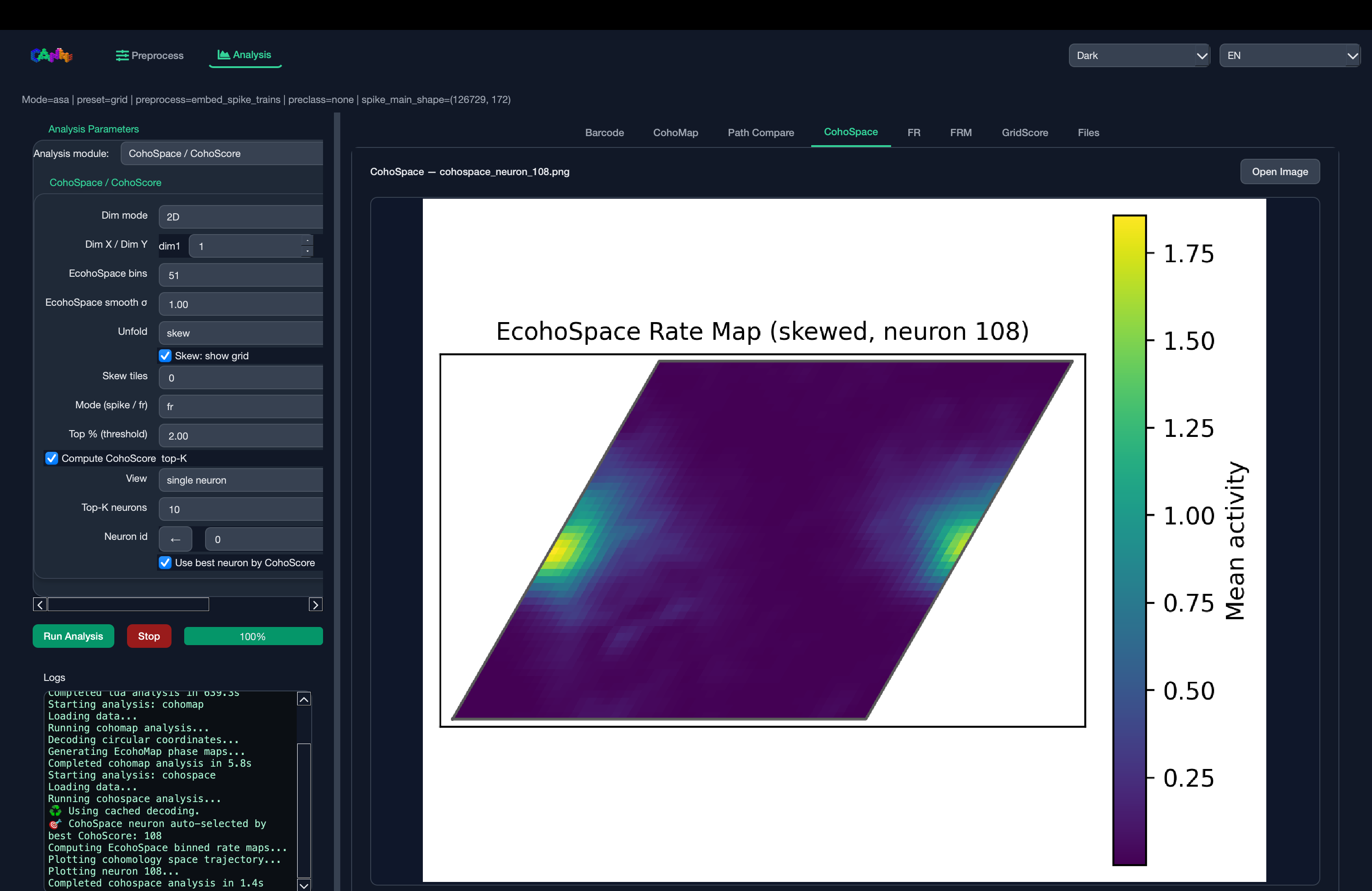}
    \caption{EcohoSpace mode.}
  \end{subfigure}
  \caption{Current ASA graphical interface.  The GUI exposes the same
  cache-aware pipeline used by the scripting interface: persistent
  homology produces barcode outputs, persistent cohomology produces
  decoded circular coordinates, EcohoMap maps these coordinates back to
  physical space, and EcohoSpace visualizes single-cell activity on the
  decoded phase manifold.}
  \label{fig:asa-gui}
\end{figure}
\FloatBarrier

ASA can be used through Python scripts, a terminal user interface,
and a \pkg{PySide6} graphical interface.  The GUI is organized around
named analysis modes: preprocessing, TDA, decoding-dependent
CohoMap/EcohoMap, CohoSpace/EcohoSpace, PathCompare, firing-rate
summary plots, firing-rate maps, and GridScore.  Modes declare their
dependencies, so analyses that require decoded circular coordinates
reuse prior TDA outputs instead of recomputing them silently.

Result management is cache-aware.  ASA stores artefacts under an
input-hash-specific results directory, with per-stage outputs such as
\texttt{persistence\_result.npz}, \texttt{barcode.png},
\texttt{cohomap\_data.npz}, \texttt{cohomap.png},
\texttt{cohospace\_data.npz}, \texttt{cohospace.png}, and
\texttt{summary.json}.  Cached stages are reused only when the
required artefacts exist and the stored parameters match the current
request.  This design is crucial for heavy analyses, because a user
can rerun visualization, cohomology decoding, or module-level
summaries without rerunning the most expensive persistent-homology
step.

This engineering layer is part of the scientific contribution of ASA.
It makes topology-based neural analysis auditable: a published figure
can be traced to the input file, preprocessing settings, TDA
parameters, shuffle count, selected cells, decoded coordinates, and
rendered artefacts.

%% file: sections/results.tex

\section{Capabilities and Reproduced Studies}
\label{sec:capabilities}

\begin{figure}[!htbp]
  \centering
  \includegraphics[width=\textwidth]{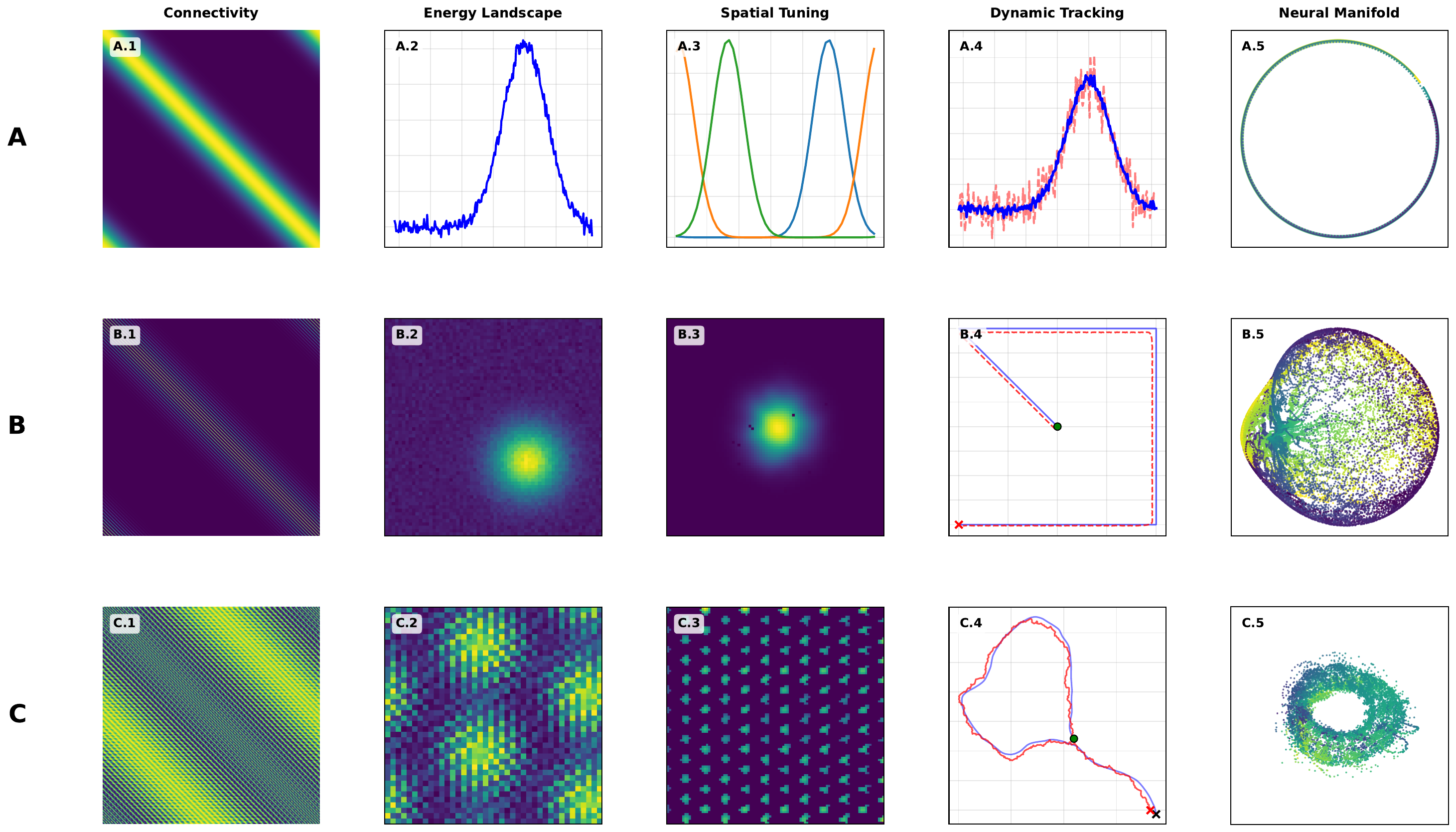}
  \caption{Three canonical CANN models side by side. The five columns
  show, from left to right, the recurrent connectivity, the
  static energy landscape, the spatial tuning of three example
  neurons, the dynamic tracking of an external input, and the
  low-dimensional neural manifold recovered from population
  activity. Row~A: 1D ring CANN (head-direction encoding). Row~B:
  2D torus CANN (place-field encoding). Row~C: 2D grid-cell
  network (path integration from velocity). The toroidal manifold
  in row~C is the topological signature recovered by persistent
  cohomology in Sec.~\ref{sec:application}.}
  \label{fig:fig4}
\end{figure}

\subsection{1D CANN: Bump Dynamics and Energy Landscape}
\label{sec:cann1d}

The simplest model in the library is a ring-shaped continuous
attractor with $N{=}512$ excitatory units and the normalised
Gaussian recurrent kernel of
Eq.~\eqref{eq:cann-dynamics},
\begin{equation}
  J(z_i - z_j) \;=\;
  \tfrac{J_0}{\sqrt{2\pi}\,a}\,
  \exp\!\bigl(-(z_i - z_j)^2 / (2a^2)\bigr),
  \label{eq:cann1d}
\end{equation}
where $z_i \in [-\pi, \pi)$ is the preferred head direction of
neuron $i$, and $J_0$ and $a$ set the strength and width of the
recurrent kernel.  The divisive-normalisation firing rate in
Eq.~\eqref{eq:cann-dynamics} implements a uniform global
inhibition set by $k$.  The script
\texttt{cann1d\_oscillatory\_tracking.py} (and the static-curve
counterpart \texttt{cann1d\_tuning\_curve.py}) drives the network with
a step input that visits four head directions in turn, $I_{\rm ext}(t) =
(z^\star(t) - z_i)$ with $z^\star(t) \in \{0,\, 0.75\pi,\, 2\pi,\,
1.75\pi\}$, and records the population rate $r(z, t)$ at $\Delta t =
0.1$~ms.

The five analytic views in row~A of Fig.~\ref{fig:fig4} together verify
that the network behaves as a head-direction ring attractor. (i)~The
connectivity plot shows a single symmetric Gaussian peak centred at
$z_i = z_j$ (no preferred direction). (ii)~The energy landscape
$E(\mathbf{r}) = -\tfrac{1}{2} \mathbf{r}^\top W \mathbf{r}
+ \mathbf{I}^\top \mathbf{r}$ (with $r_i = [u_i]_+^2/(1 +
k\sum_j [u_j]_+^2)$ from Eq.~\eqref{eq:cann-dynamics}) has a single
circular valley that respects the ring topology: $z$ and $z + 2\pi$
are the same attractor. (iii)~Tuning curves of
neurons $i \in \{128, 256, 384\}$ are well described by Gaussians of
width $\sigma_{\rm tuning} \approx 0.35$~rad, centred at their preferred
directions and with peak rates of $0.8$--$1.0$. (iv)~When the external
input jumps, the bump tracks the stimulus with a small but finite
latency set by the membrane time constant $\tau = 1\,$ms
(which is a numerical convenience; the qualitative attractor
behaviour is robust to larger $\tau$ values). (v)~The
population manifold, recovered by PCA on $r(\cdot, t)$ and projected
onto the first two principal components, is a closed loop, confirming
the ring topology at the population level.

This model is the textbook instantiation of the Amari neural
field~\cite{amari1977dynamics,wu2008dynamics} and the Wu--Amari
continuous-attractor framework~\cite{wu2016continuous}.  We reproduce
it here as a sanity check before the more elaborate models below.

\subsection{2D CANN: Spatial Encoding}
\label{sec:cann2d}

Extending Eq.~\eqref{eq:cann1d} to two dimensions, the
\texttt{CANN2D} class in \texttt{canns.models.basic.cann} places neurons on a
$100 \times 100$ sheet and uses the isotropic 2D Gaussian recurrent
kernel of Eq.~\eqref{eq:cann-dynamics},
\begin{equation}
  J(\mathbf{z}_i - \mathbf{z}_j) \;=\;
  \tfrac{J_0}{2\pi\,a^2}\,
  \exp\!\bigl(-\|\mathbf{z}_i - \mathbf{z}_j\|^2 / (2a^2)\bigr),
  \label{eq:cann2d}
\end{equation}
where $\mathbf{z}_i \in [-\pi, \pi)^2$ is the preferred 2D
stimulus of neuron $i$, and $J_0$ and $a$ set the strength and
width of the recurrent kernel.  Combined with the same
divisive-normalisation firing rate
$r_i = [u_i]_+^2/(1 + k\sum_j [u_j]_+^2)$ from
Eq.~\eqref{eq:cann-dynamics}, this produces a bump on a 2D flat
torus $\mathbb{T}^2$.  The script \texttt{cann2d\_tracking.py}
drives the network with a piecewise-constant spatial input that
visits six 2D way-points on a $3 \times 3$ m environment, then
records the membrane potential $u(\mathbf{z}, t)$ for offline
analysis.

Row~B of Fig.~\ref{fig:fig4} shows that the same five-views framework
applies with little modification. The connectivity is still a single
Gaussian (isotropic in the plane). The energy landscape now has a
single circular valley in the $z_y$ direction at each fixed $z_x$ (a
2D version of row~A.~2), so the static analysis reduces to the 1D
case slice by slice. Spatial tuning curves are well described by 2D
Gaussians, and three example neurons tile the environment evenly.
Dynamic tracking shows the 2D bump following a piecewise-linear
trajectory with a comparable membrane-time-constant latency. The
population manifold is a single 2D closed surface (visually a
rounded square), confirming the toroidal topology of the 2D CANN
state space. This model is the standard place-field attractor
analysed in~\cite{fung2010moving,wu2016continuous}.

\subsection{Grid Cell Network: Toroidal Manifold}
\label{sec:gridcell}

The 2D CANN above has a single bump and a 2D-torus state space. The
grid cell network, by contrast, is constructed to have a periodic
hexagonal array of bumps, and its state space is the same torus
$\mathbb{T}^2$ but tiled many times by the bump pattern. The library
exposes two flavours: \texttt{GridCell2DPosition} (direct position
input) and \texttt{GridCell2DVelocity} (path-integration from
velocity). The path-integration flavour follows
Burak and Fiete~\cite{burak2009accurate} and is verified by
\texttt{grid\_cell\_velocity\_path\_integration.py}, which runs the
network on a 10~s, $2.2 \times 2.2$~m open-loop trajectory and
reconstructs position by tracking the centre of mass of the bump
lattice.

The verification script reports $R^2 > 0.99$ between the reconstructed
trajectory and the ground truth, mean position error
$\bar{e} \approx 0.05$~m, and proportional scaling factor
$\approx 1.0$ (cf.\ the docstring of
\texttt{grid\_cell\_velocity\_path\_integration.py}, lines
115--125). The complementary script
\texttt{grid\_cell\_velocity\_spatial\_analysis.py} systematically
samples the environment to compute firing-rate maps and reports
grid scores $> 0.6$ with clear hexagonal spatial autocorrelations.

Row~C of Fig.~\ref{fig:fig4} summarises the analysis. The
connectivity is a sum of three Gaussian components, each shifted by
one-third of the lattice constant along the canonical 60$^\circ$ axis,
which is the construction that produces the hexagonal lattice. The
energy landscape is a 2D lattice of identical wells arranged in a
hexagonal pattern, and the static bumps therefore live at the
lattice points. The spatial tuning of three example cells now
consists of three regularly spaced Gaussian bumps in the same
hexagonal arrangement. Dynamic tracking shows the entire lattice
translating as a rigid body when the input velocity is constant, with
a small phase lag. The population manifold is the same torus
$\mathbb{T}^2$ as in the 2D CANN, but each cycle of the torus
corresponds to a full lattice translation; this is the type of
toroidal signature reported by Gardner et
al.~\cite{Gardner2022Toroidal} and the topological signal that
ASA tests in real MEC grid-cell modules in
Sec.~\ref{sec:application}.

\subsection{Spike-Frequency Adaptation: Anticipative Tracking}
\label{sec:sfa}

Adding a slow adaptation variable $a_i$ on top of the 1D
CANN, with dynamics
\begin{equation}
  \tau_a \, \dot{a}_i \;=\; -a_i + m \, u_i,
  \label{eq:sfa}
\end{equation}
and entering the membrane equation of Eq.~\eqref{eq:cann-dynamics}
as an additive negative feedback $-a_i$ (replacing the
$\tau\,\dot u_i = -u_i + \sum_j W_{ij}\,r_j + I_i$ line by
$\tau\,\dot u_i = -u_i + \sum_j W_{ij}\,r_j - a_i + I_i$), gives
the SFA-CANN that we re-implement from
Mi et al.~\cite{mi2014spike} and the more recent analyses of
Li et al.~\cite{li2025dynamics}.  (Both
of these authors are co-authors of the present report; the
\texttt{canns} re-implementation is the version we ship.)  The
script \texttt{cann1d\_oscillatory\_tracking.py} (run with a
single moving input rather than a step sequence) shows that the
SFA bump no longer lags the input: at moderate input speeds the
population bump anticipates the stimulus position by a phase
offset that grows linearly with speed before saturating, exactly
as predicted in \cite{mi2014spike}. The same anticipation
regime is visible in the population manifold: the manifold
deforms into a travelling-wave limit cycle whose phase leads the
input phase. In the oscillatory-tracking case, this manifests as
anticipative theta-phase sweeps that precess faster than the
input, and at higher speeds the bump emits a single forward
``sweep'' per theta cycle, matching the phase-precession
phenomenology reported for hippocampal place
cells~\cite{o1993phase,chu2024firing}.

The SFA time constant $\tau_a$ is the only free parameter that
controls the strength of anticipation. The default setting in
\texttt{cann1d\_oscillatory\_tracking.py} is $\tau_a = 100$~ms, which
is biologically motivated by the calcium-activated potassium
conductance that underlies SFA in cortical and hippocampal
pyramidal neurons, and which produces a comparable time constant
\textit{in vitro}. With $\tau_a = 100$~ms and a sinusoidal input of
period $200$~ms the bump leads the input by approximately
$30^\circ$ of phase, again consistent with the original
Mi~et~al.~result~\cite{mi2014spike}. The script exposes
$\tau_a$ and the input speed as keyword arguments, so users can
sweep both and inspect the resulting phase-lead curve without
modifying the model code. This is the first example in the
library in which a \emph{temporal} neural computation (anticipation)
emerges from a \emph{spatial} attractor model with a single
additional slow variable, and it is a useful template for users who
want to add their own slow variables to the 1D or 2D CANN.

\subsection{Theta-Sweep: HD / Place / Grid Cell Systems}
\label{sec:thetasweep}

The theta-sweep family of models (\code{DirectionCellNetwork},
\code{PlaceCellNetwork}, \code{GridCellNetwork} in
\texttt{canns.models.basic.theta\_sweep\_model}) is implemented in the
same canns library and shares the same theta-modulation back-end:
the input drive to each cell is multiplied by
$1 + \alpha \, v \, \cos(2\pi f_\theta t + \phi)$, with $v$ the
instantaneous speed (and a similar term proportional to the
angular-velocity gain for direction cells), and the network is
otherwise a 1D or 2D CANN. This back-end, originally described by
O'Keefe and Burgess~\cite{okeefe1996geometric}, is the substrate of the recent phase
precession work of Ji et al.~\cite{ji2025phase,ji2025systems} and
the place-cell sweep model of Chu et al.~\cite{chu2024firing},
both of which are re-implemented in \texttt{canns} by their
original authors (who are co-authors of the present report).

The three example scripts in
\texttt{examples/cann/}
(\texttt{theta\_sweep\_grid\_cell\_network.py},
\texttt{theta\_sweep\_place\_cell\_network.py},
and the direction-cell example in the same directory) drive the
corresponding network with an open-loop navigation trajectory and
record the bump position as a function of the theta phase. The
recovered ``internal trajectory'' plotted on top of the real
trajectory shows a saw-tooth pattern, the same alternating
sweep signature as in the original publications. The grid-cell
theta-sweep example additionally produces a 3D rendering of the
twisted-torus manifold, with the internal trajectory colouring the
torus as the animal moves. The place-cell example uses the T-maze
geometry of~\cite{chu2024firing}
(\texttt{TMaze\-RecessOpen\-Loop\-Navigation\-Task}
with $w=0.84\,$m, $l_s=3.64\,$m, $l_{\rm arm}=2.36\,$m) and shows
phase precession at the T-junction as the agent slows down.

\subsection{Hierarchical Path Integration}
\label{sec:hierarchical}

The \texttt{HierarchicalNetwork} model implements a multi-scale
band/grid/place hierarchy that re-implements the architecture
of~\cite{chu2025localized} (the original author is a co-author
of the present report) and draws on the broader path-integration
literature~\cite{mcnaughton2006path,
samsonovich1997path,etienne2004path}. The default configuration has
$5$ modules, each with $30$ place cells, and a 2D velocity input is
fed simultaneously to all modules. Each module contains its own band
cells in the $x$ and $y$ directions and its own grid-cell lattice,
and place cells are read out as the conjunction of all upstream
band/grid inputs.

The example \texttt{hierarchical\_path\_integration.py} runs the
network on a 50\,000-step, $5 \times 5$~m trajectory and computes
the firing-rate heat maps for all four populations. The expected
qualitative results are: (i)~band cells have one or two oriented
stripes per module, with the stripe orientation rotating across
modules; (ii)~grid cells within a module share a common lattice but
differ in phase, and across modules the lattice scale changes;
(iii)~place cells have a single Gaussian field per cell, and the
field sizes grow with the module index (multi-scale
representation). The combination of these three properties is the
prediction of~\cite{chu2025localized} that localized phase coding
and space coding \emph{complement} each other to form a robust
multi-scale representation, and the canns library provides a
single API for all three populations so the result can be inspected
side by side.  The computed firing-rate heat maps and the trajectory
visualization for this model are produced by the script
\texttt{examples/cann/hierarchical\_path\_integration.py} and explained
in the companion tutorial
\texttt{docs/en/3\_full\_detail\_tutorials/01\_cann\_modeling/06\_hierarchical\_network.ipynb}
in the \texttt{canns} repository.

\subsection{Brain-Inspired Learning Examples}
\label{sec:braininspired}

The \texttt{examples/brain\_inspired/} directory is the
\texttt{trainer} section of the library, demonstrating classic
learning rules on simple models rather than on full CANN
architectures. These examples are the recommended entry point for
users who want to understand how a CANN's recurrent weights can be
\emph{learned} (rather than hand-tuned) before adding the learning
rule to a full spatial-coding model. Each example is a single
self-contained script that runs end-to-end in a few minutes and
produces a single figure under
\texttt{examples/outputs/brain\_inspired/}.

\textbf{Amari--Hopfield associative memory.}
\texttt{hopfield\_train.py} instantiates an
\texttt{Amari\-Hopfield\-Network} (discrete mode, $\{{\pm}1\}$
states) and trains it with a \texttt{Hebbian\-Trainer} on four
$128 \times 128$ binary images pre-processed from
\texttt{skimage.data}. The trained network is then queried with
$30\%$ corrupted versions of the four patterns. Hebbian
learning, originally formulated
in~\cite{hebb2005organization,amari1977neural,hopfield1982neural},
recovers the four stored patterns from these corrupted cues, with
overlap $> 0.9$ for the cleanest images. The companion
\texttt{hopfield\_energy\_diagnostics.py} adds an
\texttt{Hopfield\-Analyzer} that estimates the storage capacity
$\approx N / (4 \ln N)$, computes the energy of each pattern,
and quantifies the recall quality as a function of the
corruption level. Together, these two examples illustrate the
\emph{separation of concerns} design of canns: the
\texttt{Hebbian\-Trainer} updates the weights, the
\texttt{Amari\-Hopfield\-Network} runs the dynamics, and the
\texttt{Hopfield\-Analyzer} provides diagnostics, with no
component reaching into the internals of another.

\textbf{Oja vs.\ Sanger PCA.}
\texttt{oja\_vs\_sanger\_comparison.py} trains a 3-neuron
\texttt{LinearLayer} on synthetic 50-dimensional data with three
known principal components and compares Oja's rule
\cite{oja1982simplified} with Sanger's rule
\cite{sanger1989optimal}. As predicted theoretically, all three Oja
neurons converge to the first principal direction and only the
first one aligns with the true PC (cosine similarity $\approx
0.96$, $\approx 0.45$, $\approx 0.30$); the three Sanger neurons
align with all three PCs (cosine similarity $\approx 0.96$,
$0.92$, $0.88$) and the inter-component dot product drops below
$0.1$. This is the cleanest demonstration in the library that
\emph{principal component extraction with multiple neurons requires
Gram--Schmidt-style orthogonalisation}, a point often glossed over
in textbook treatments.

\textbf{STDP temporal learning.}
\texttt{stdp\_temporal\_learning.py} trains a 5-output, 20-input
spiking layer with a JIT-compiled \texttt{STDPTrainer} on temporal
Poisson spike patterns structured into three temporal groups
(early, middle, late) of five inputs each, plus 5 noise
inputs~\cite{bi1998synaptic}. After 20 epochs the early-spiking
inputs develop positive weight changes (LTP), the late-spiking
inputs develop negative or near-zero changes (LTD), and the noise
inputs remain at their initial values. The script prints the per-group
weight changes and shows weight-evolution trajectories and spike
rasters, making STDP's temporal credit assignment directly visible
without resorting to a black-box simulator.

\section{Application: Analyzing Real Neural Recordings with ASA}
\label{sec:application}

This section presents ASA as a general attractor-topology analysis
system rather than a torus-only grid-cell script.  The primary
evidence in this section is organized around real neural recordings
and real grid-cell modules.  Simulated CANN/RNN datasets are used only
as controlled benchmarks, because their expected topology is known or
strongly constrained.  The examples below therefore include both
positive recoveries and diagnostic analyses that expose partial or
unstable topology.

\subsection{Ring and Torus Analyses in ASA}
\label{sec:asa-ring-torus-results}

ASA can decode either one or two circular coordinates depending on
the topology supported by the persistent $H_1$ generators.  A single
stable $H_1$ feature supports a ring-like analysis, appropriate for
head-direction cells, band cells, and 1D CANNs.  Two stable
$H_1$ features support a toroidal analysis, appropriate for grid-cell
modules.  This distinction is important because the same GUI modules
--- EcohoMap and PathCompare --- can operate in one-dimensional or
two-dimensional mode.  The one-dimensional mode should be presented
before the toroidal examples so that readers do not infer that ASA is
restricted to $T^2$.

\begin{figure}[t]
  \centering
  \begin{subfigure}[t]{0.32\textwidth}
    \includegraphics[width=\textwidth]{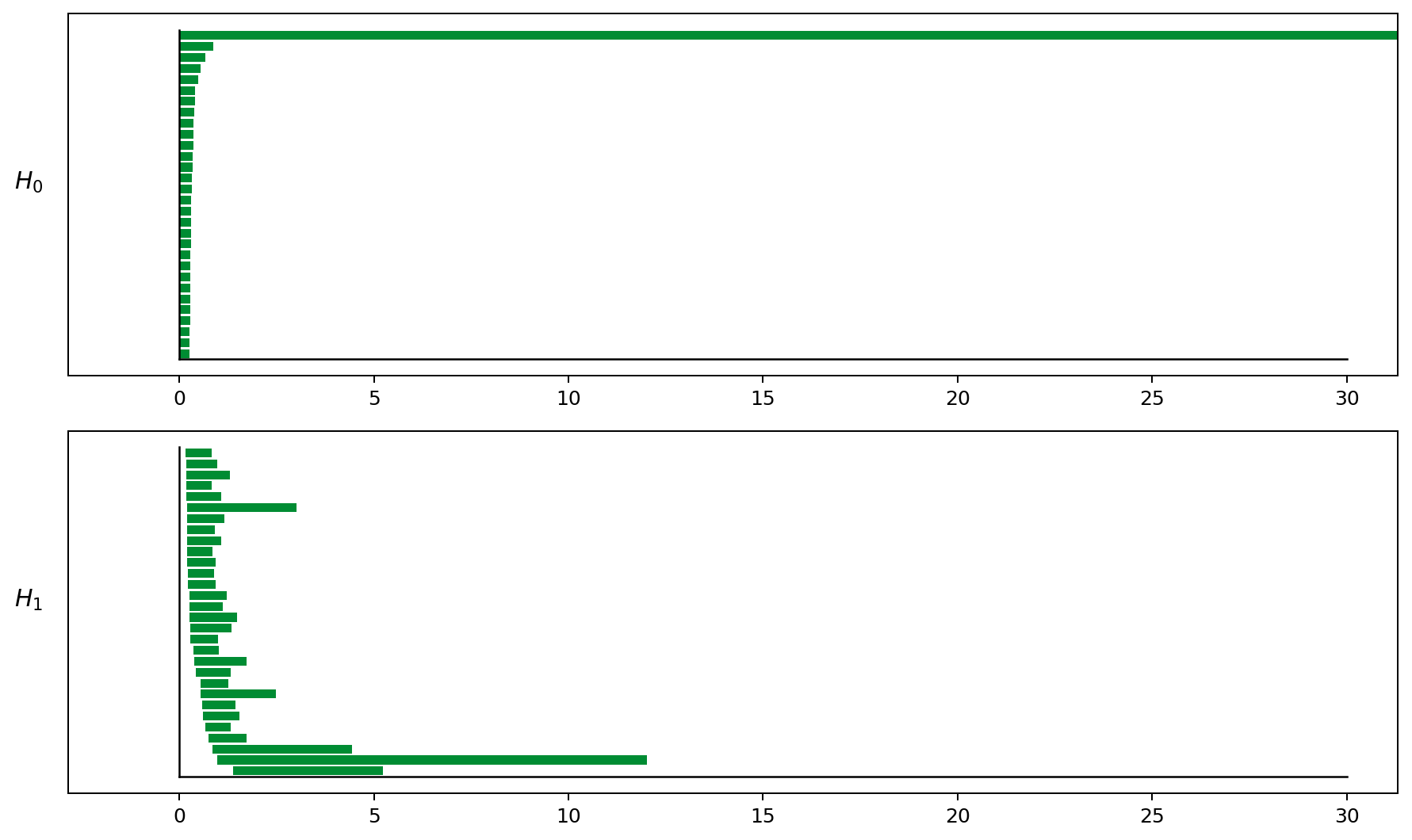}
    \caption{One dominant $H_1$ barcode.}
  \end{subfigure}
  \hfill
  \begin{subfigure}[t]{0.64\textwidth}
    \includegraphics[width=\textwidth]{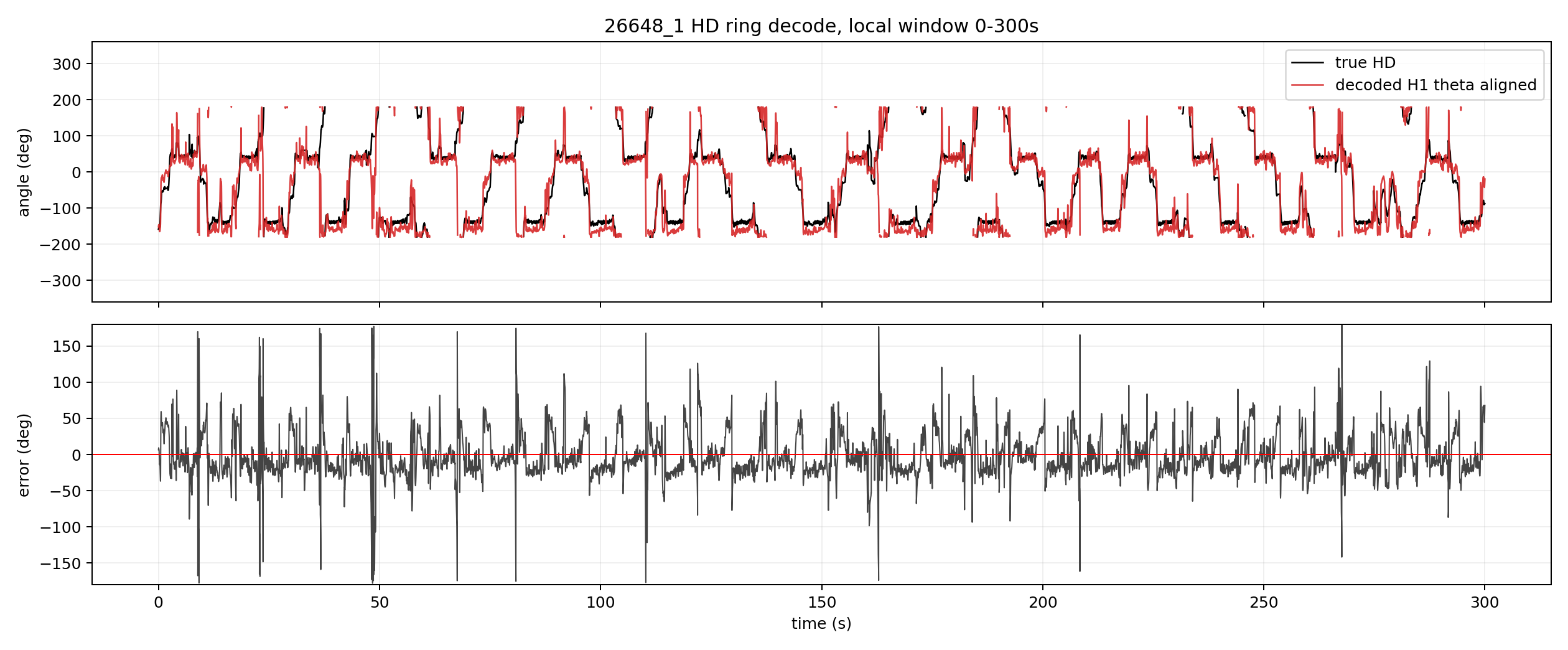}
    \caption{Decoded phase aligned to true head direction.}
  \end{subfigure}
  \caption{ASA analysis of a real MEC head-direction-cell subset from
  session 26648\_1.  MEC units were selected from the full MEC
  recording using head-direction tuning strength
  (\texttt{rmc.hd.mvl}$>0.5$) while excluding grid cells.  The barcode
  contains one dominant $H_1$ feature, consistent with a ring-like
  state space.  The single cohomology coordinate was then aligned by a
  global sign and phase offset to the recorded head direction.  The
  aligned coordinate follows the behavioural head-direction trace with
  a mean absolute circular error of approximately $23.8^\circ$.}
  \label{fig:asa-hd-ring}
\end{figure}
\FloatBarrier

\subsection{Real MEC Grid-Cell Modules Show Heterogeneous Topology}
\label{sec:asa-real-mec-heterogeneous}

The main toroidal application should be based on real MEC grid-cell
modules rather than on virtual controls.  We therefore use the
Left--Right MEC module collection as the central dataset for this
section.  The first-pass screen contains 66 module-level datasets,
with module sizes ranging from 13 to 215 cells and recording lengths
from 50,000 to 500,000 samples.  Each module is analysed with the
time-indexed ASA route: spike/rate preprocessing, standardization,
PCA to six dimensions, persistent homology, and cohomology decoding
when stable $H_1$ features are present.

The key diagnostic analysis is a real-data topology-quality summary.
For each module ASA records the top $H_1$ lifetimes, the shuffle null
percentiles, and whether the first and second $H_1$ bars exceed the
chosen null threshold.  This allows real MEC modules to be grouped
conceptually into three classes:
\begin{align}
  \text{strong torus:} \quad&
  H_1^{(1)}>q,\quad H_1^{(2)}>q,\\
  \text{partial/ring-like:} \quad&
  H_1^{(1)}>q,\quad H_1^{(2)}\le q,\\
  \text{weak:} \quad&
  H_1^{(1)}\le q,
\end{align}
where $q$ is a pre-specified shuffle percentile.  This real-data
classification is the appropriate place to show that ASA does not
force every grid-like dataset into a toroidal interpretation: some
real modules show two stable cycles, some only one, and some fail the
shuffle criterion.  The broad screen was therefore used both as a
quality-control step and as the source of the high-quality cohort in
Sec.~\ref{sec:asa-selected20}, where strict maxdim-2 shuffle controls
are shown explicitly.

\subsection{Cohomology Decoding Links Topology to Physical Space}
\label{sec:asa-real-cohomap}

Persistent homology alone only reports candidate holes in the point
cloud.  ASA's next step is to decode persistent cohomology generators
and test whether the resulting circular coordinates are meaningful in
the behavioural or physical coordinate system.  For ring-like data,
this means checking whether a single decoded phase tracks a
one-dimensional variable such as head direction, band phase, or ring
position.  For grid-cell modules, this means checking whether two
decoded phases form interpretable EcohoMaps over the animal's
physical environment.

\begin{figure}[!htbp]
  \centering
  \includegraphics[width=\textwidth]{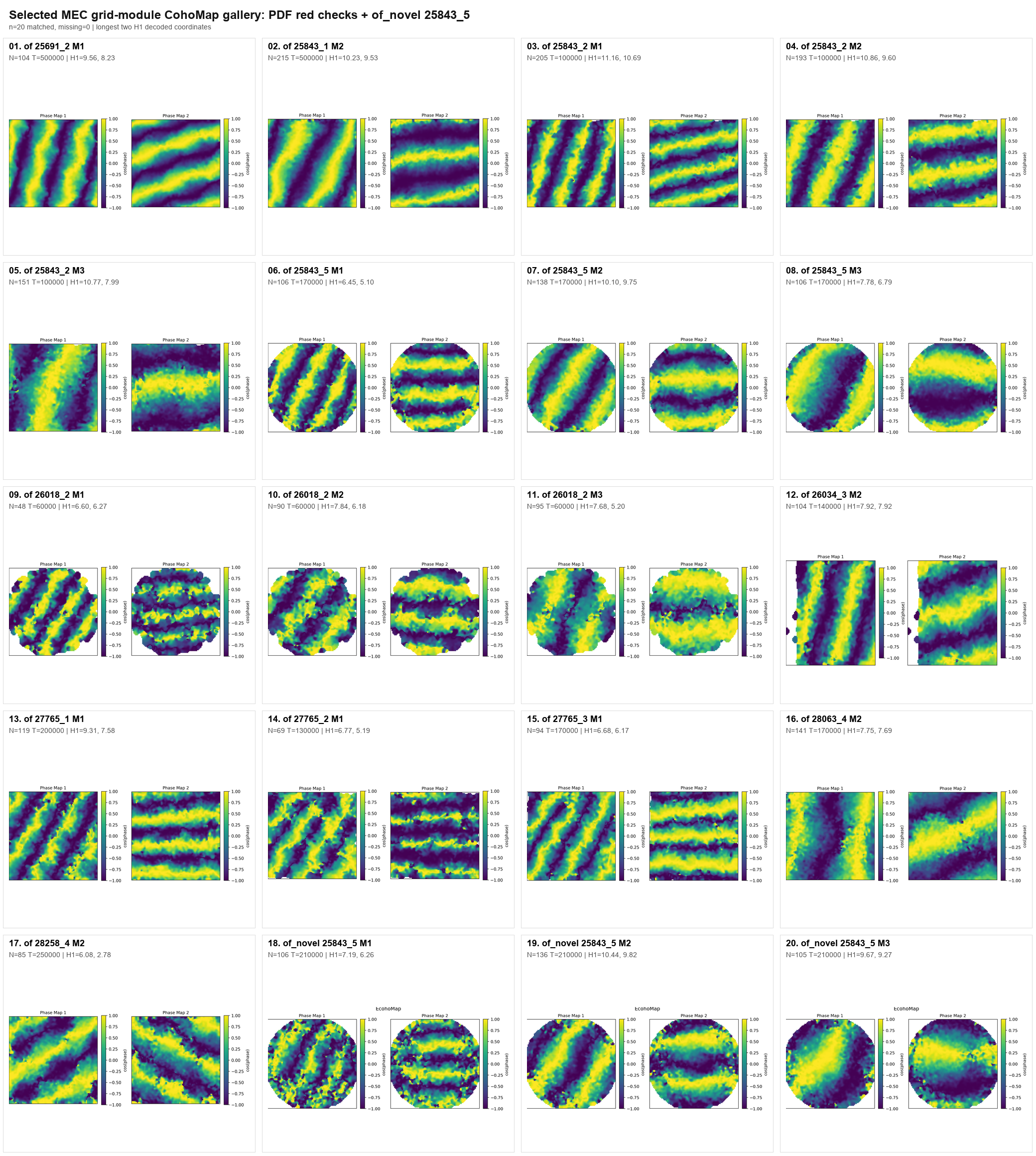}
  \caption{EcohoMap gallery for selected real MEC grid modules.  ASA
  decodes two persistent cohomology coordinates and maps their circular
  means back to the animal's physical environment.  The resulting phase
  maps provide a spatial sanity check for the abstract $H_1$ generators
  recovered from the population point cloud.}
  \label{fig:asa-selected20-cohomap}
\end{figure}
\FloatBarrier

\begin{figure}[!htbp]
  \centering
  \includegraphics[width=\textwidth]{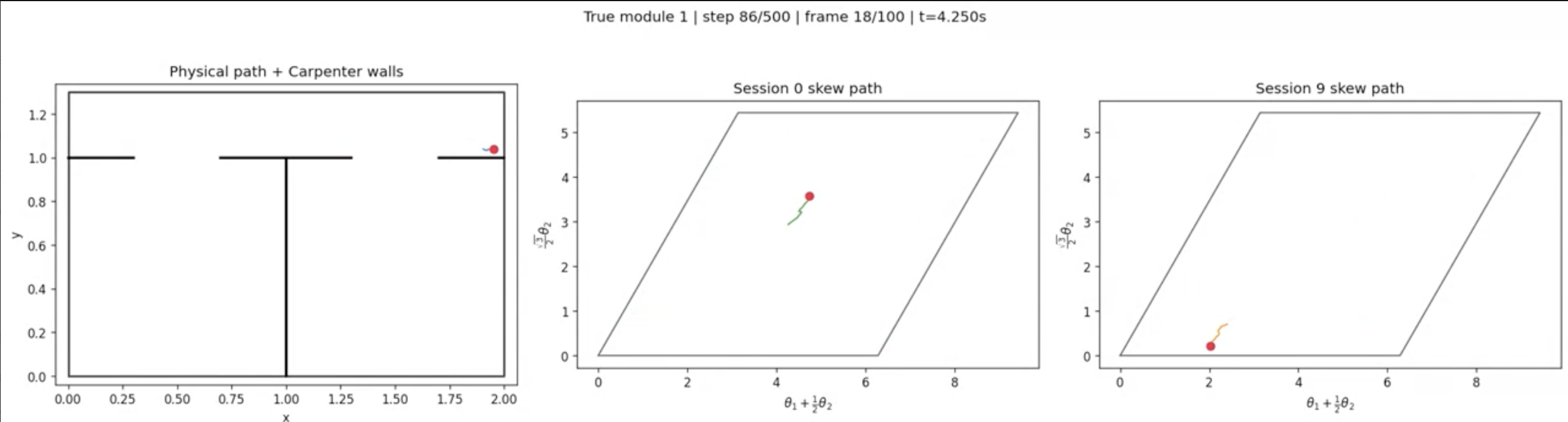}
  \caption{PathCompare visualization for a grid module across two
  sessions.  The left panel shows the physical path in the Carpenter
  environment, while the middle and right panels show the corresponding
  decoded skew-torus trajectories for sessions~0 and~9.  This
  visualization tests whether cohomology coordinates preserve
  behaviourally meaningful trajectory structure.}
  \label{fig:asa-grid-pathcompare}
\end{figure}
\FloatBarrier

\subsection{Selected High-Quality MEC Modules Recover Strict Toroidal Signatures}
\label{sec:asa-selected20}

After the broad real-module screen, we use a selected high-quality
cohort to illustrate the upper end of ASA's toroidal recovery.  These
20 MEC grid-cell modules were selected because their barcodes and
CohoMap results were already visually and quantitatively promising.
They should therefore be interpreted as a positive cohort, not as an
unbiased estimate of toroidal prevalence across all MEC modules.

\begin{figure}[!htbp]
  \centering
  \includegraphics[width=\textwidth]{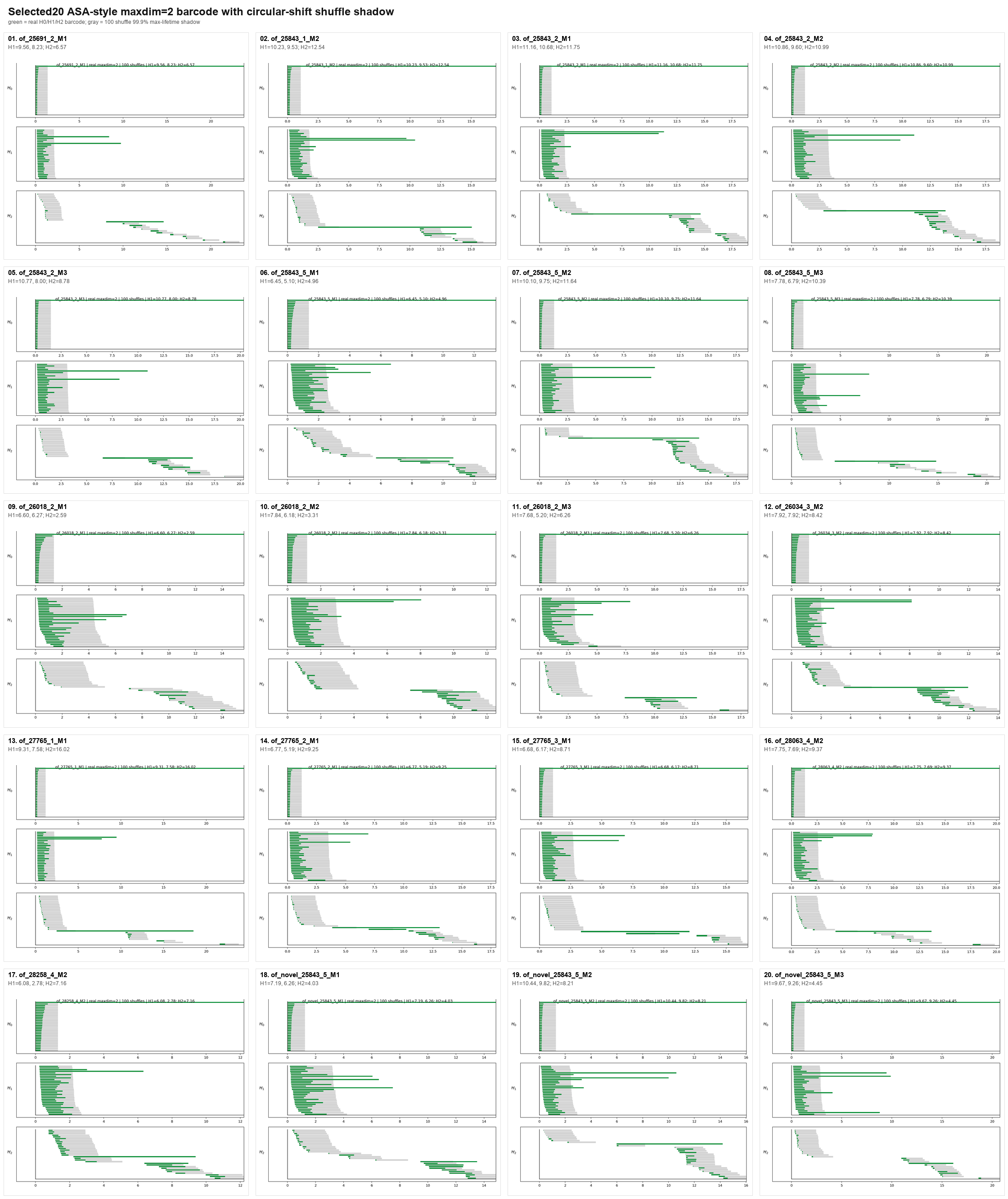}
  \caption{Twenty selected high-quality MEC grid-cell modules analysed
  with \texttt{maxdim=2} and 100 circular-shift shuffles per module.
  Green bars show real H$_0$/H$_1$/H$_2$ persistence intervals; grey
  shadows show the shuffle-derived null envelope.  This figure
  demonstrates the strict toroidal signature recoverable in favourable
  real MEC modules, but it is not a prevalence estimate.}
  \label{fig:asa-selected20-gallery}
\end{figure}

For this cohort, ASA was run with \texttt{maxdim=2} and 100
circular-shift shuffles per module.  Across the 20 modules, the
median first and second $H_1$ lifetimes were $7.883$ and $7.633$,
respectively, and the median leading $H_2$ lifetime was $8.568$.
The ASA-style barcode gallery in Fig.~\ref{fig:asa-selected20-gallery}
shows real H$_0$/H$_1$/H$_2$ intervals together with the
shuffle-derived grey null envelope.

\FloatBarrier

\subsection{GridScore, CohoScore, and Automatic Cell Selection}
\label{sec:asa-gridscore-cohoscore-results}

ASA also connects population topology to single-cell structure.
GridScore measures spatial periodicity in the firing-rate map, while
CohoScore measures concentration in the decoded cohomology phase
space.  Because lower CohoScore indicates stronger phase
concentration, a strong grid cell is expected to have high GridScore
and low CohoScore.

Across the 20 selected MEC modules, 2,207 neurons had finite scores
in both metrics.  The Pearson correlation between GridScore and
CohoScore was negative in all 20 modules.  The module-wise mean
Pearson correlation was approximately $-0.508$, the median was
approximately $-0.546$, and a Fisher-$z$ one-sided test gave
$p\approx 1.09\times 10^{-8}$.

The automatic grid-threshold workflow is one of ASA's clearest
engineering contributions.  Rather than asking the user to manually
guess a GridScore cutoff, ASA searches over top-$k$ candidate cell
sets and evaluates topology strength, shuffle stability, decoding
quality, and the GridScore--CohoScore relationship.

\begin{figure}[!htbp]
  \centering
  \includegraphics[width=0.82\textwidth]{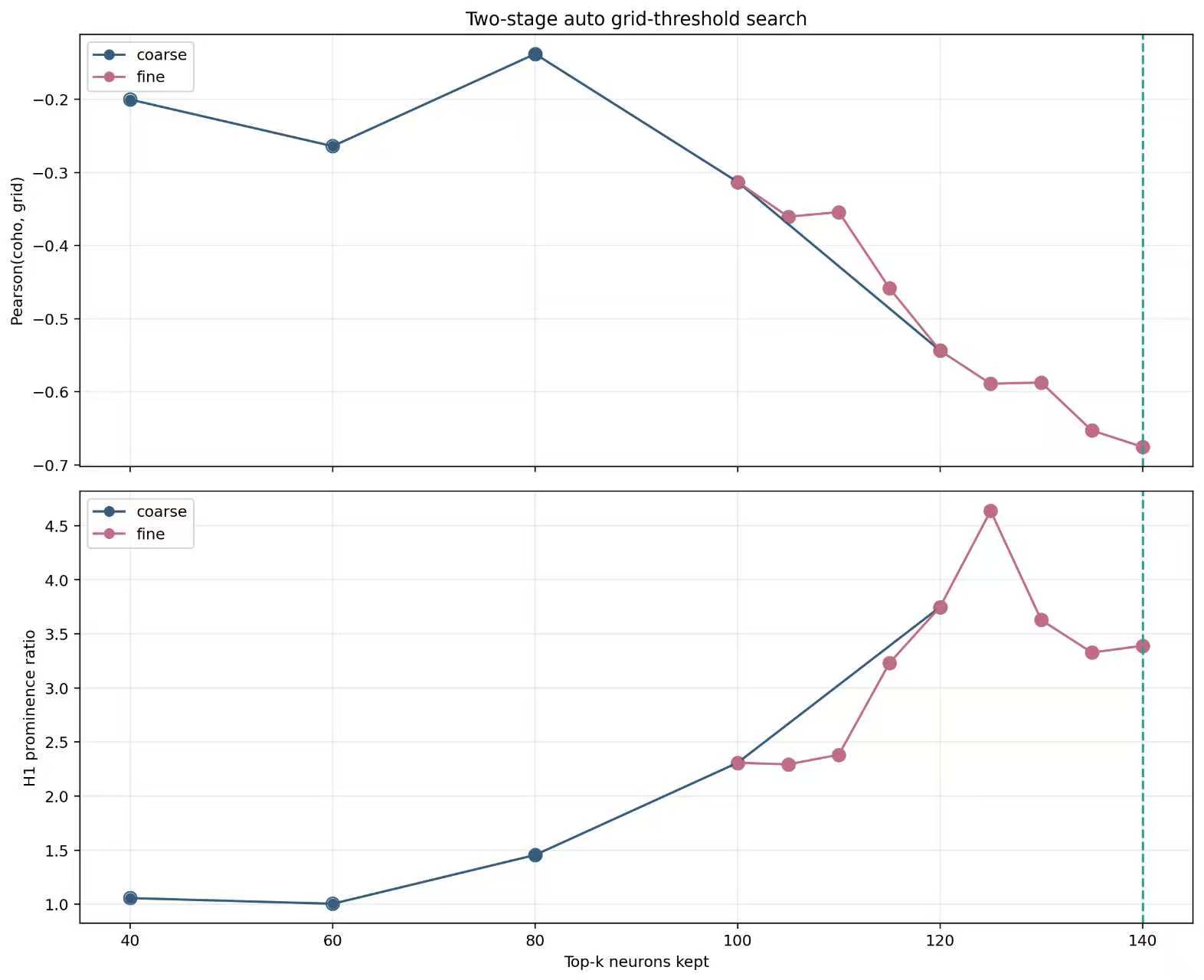}
  \caption{Two-stage automatic GridScore-threshold search.  ASA first
  performs a coarse sweep over candidate top-$k$ cell sets and then
  refines the promising region.  The upper panel tracks the
  GridScore--CohoScore correlation, while the lower panel tracks the
  relative prominence of the leading $H_1$ features.  The selected
  cutoff balances classical single-cell gridness with
  topology-aware population structure.}
  \label{fig:asa-auto-threshold}
\end{figure}
\FloatBarrier

\subsection{Simulation and Spatial-TDA Benchmarks}
\label{sec:asa-simulation-benchmarks}

Simulation results are useful for validating the ASA pipeline because
the expected topology is known or constrained by construction.  They
are not the primary evidence for ASA's real-data utility, but they
show that the same ASA interface can process model-generated dense
activity matrices.  We therefore treat virtual and RNN datasets as
controlled calibration experiments, while keeping the main-text figures
focused on real MEC and HD-cell analyses.

\begin{figure}[!htbp]
  \centering
  \begin{subfigure}[t]{0.32\textwidth}
    \includegraphics[width=\textwidth]{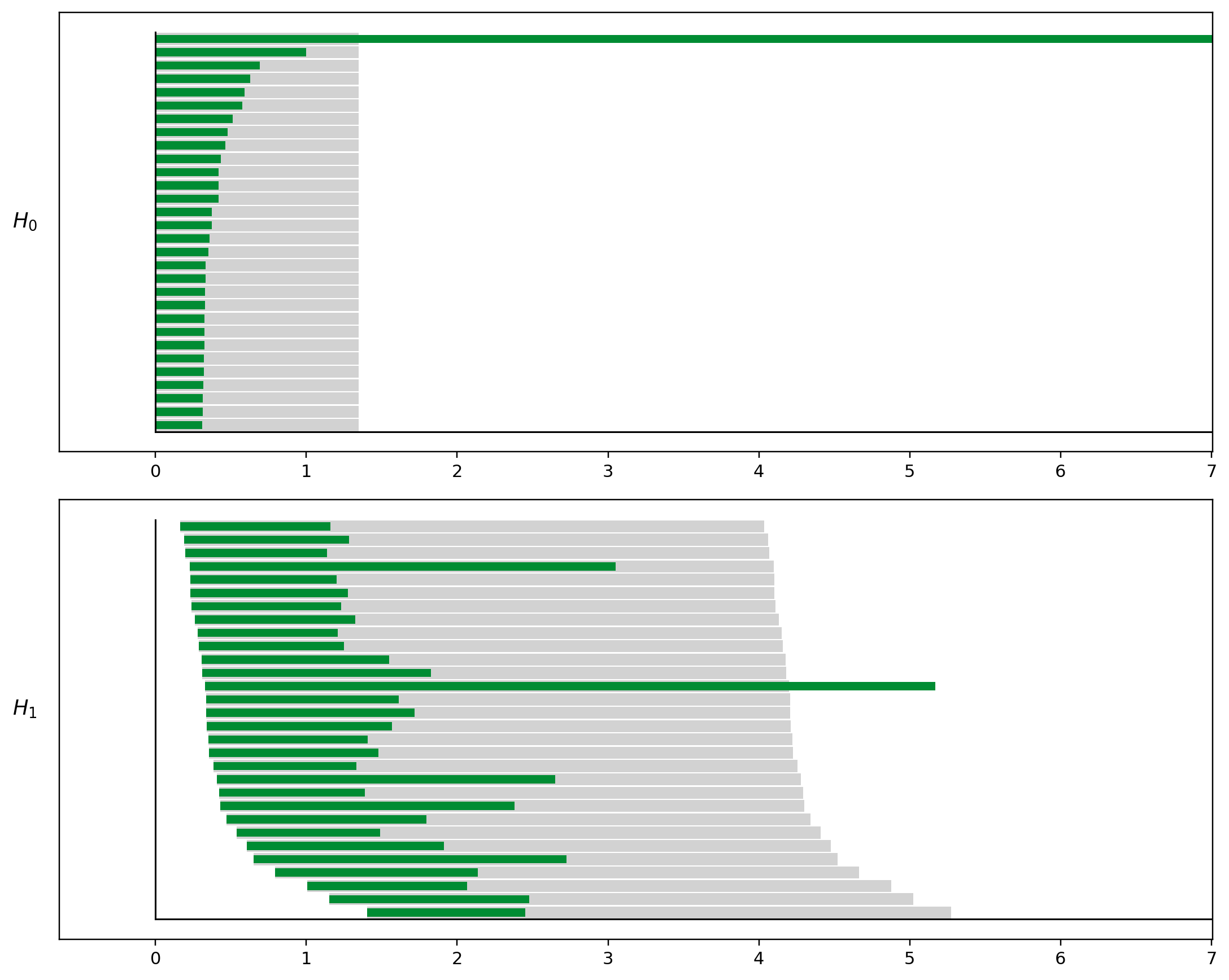}
    \caption{\texttt{764\_grid}: one significant $H_1$ feature.}
  \end{subfigure}
  \hfill
  \begin{subfigure}[t]{0.32\textwidth}
    \includegraphics[width=\textwidth]{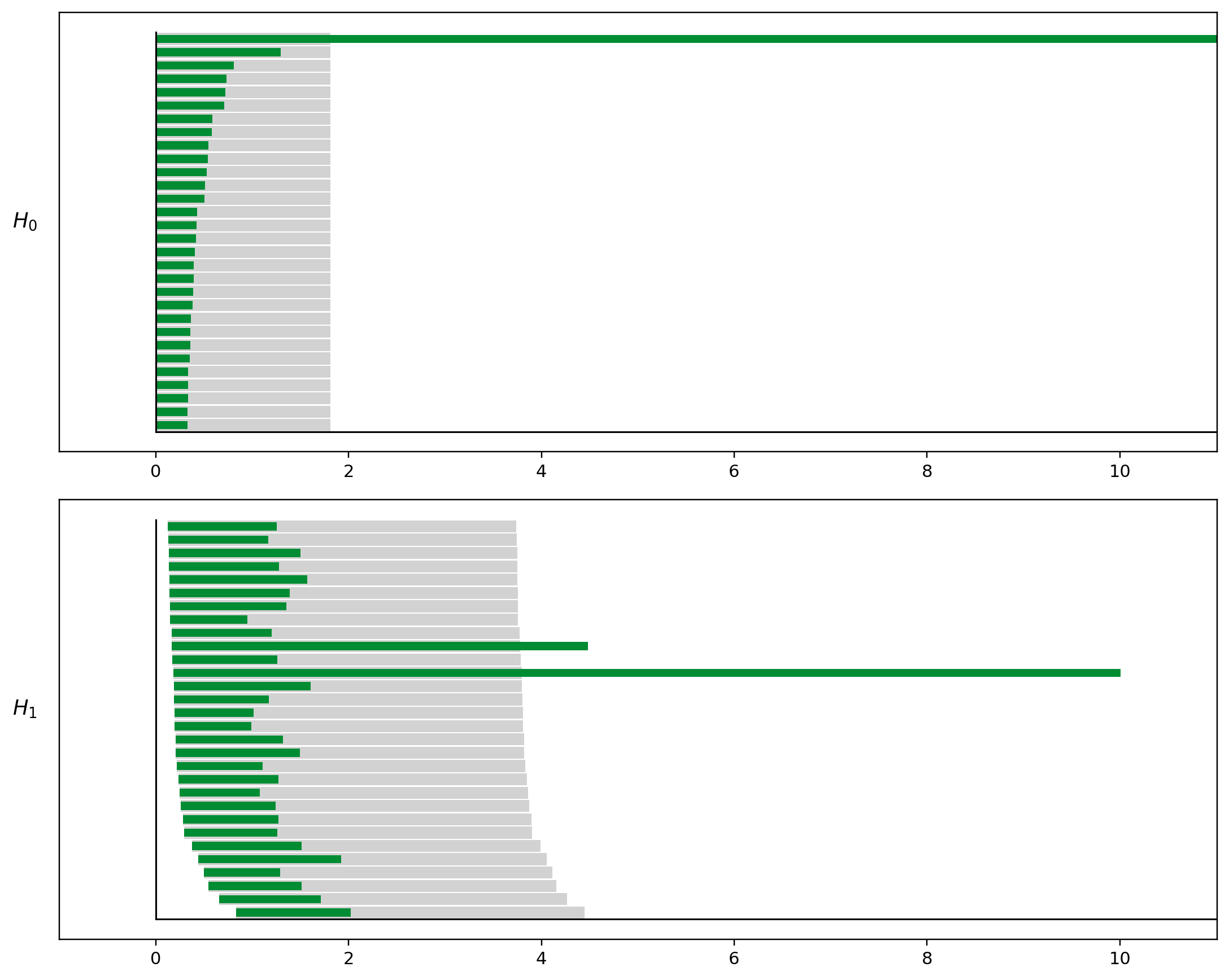}
    \caption{Virtual 4096 top764: two significant $H_1$ features.}
  \end{subfigure}
  \hfill
  \begin{subfigure}[t]{0.32\textwidth}
    \includegraphics[width=\textwidth]{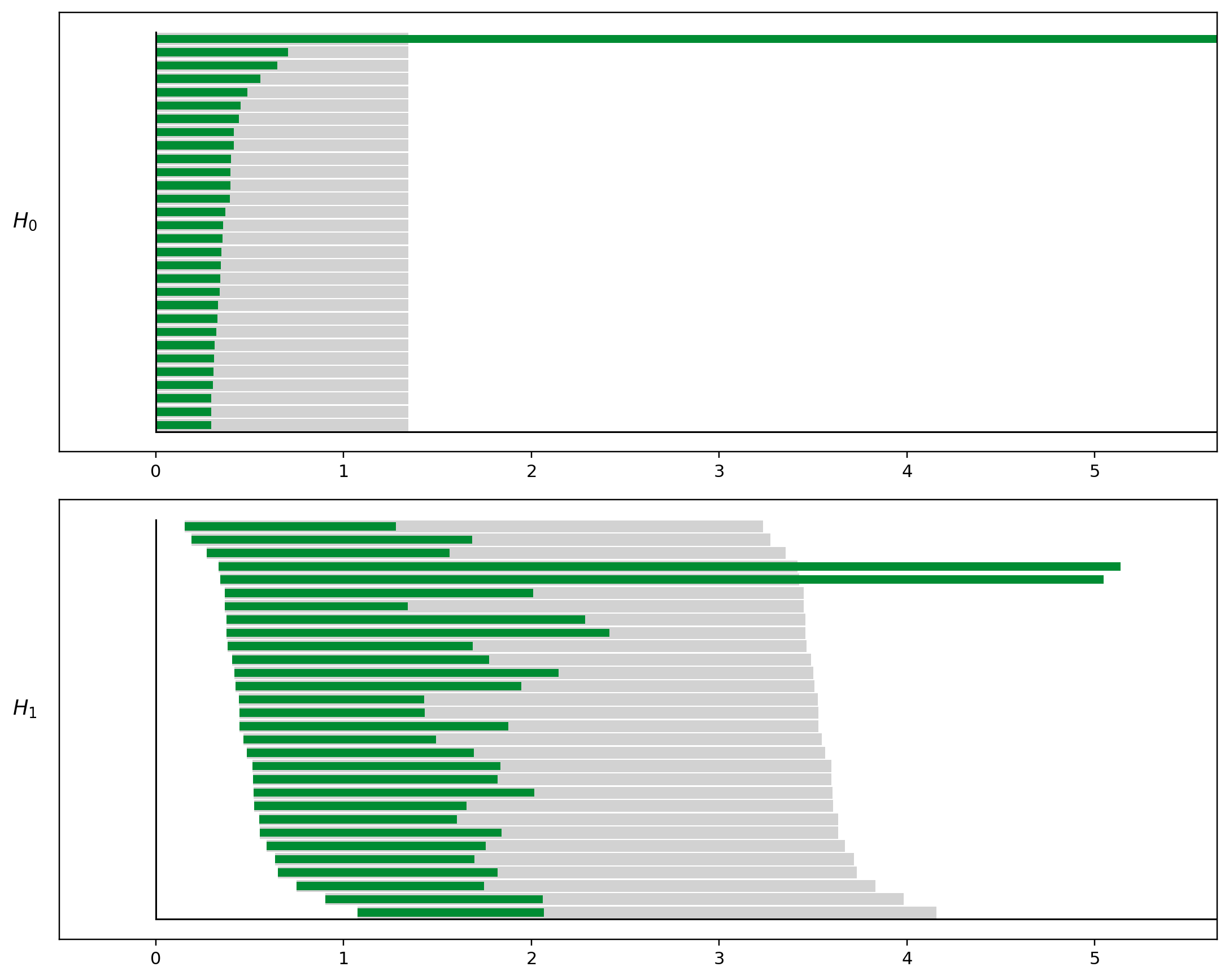}
    \caption{Vanilla RNN top764: two significant $H_1$ features.}
  \end{subfigure}
  \caption{ASA calibration runs.  Green bars show real persistence
  intervals and the grey envelope summarizes circular-shift shuffle
  controls.  The comparison demonstrates that ASA can distinguish a
  partially grid-like dataset with only one strict $H_1$ feature from
  datasets with two robust toroidal generators.}
  \label{fig:asa-calibration}
\end{figure}
\FloatBarrier

\subsection{Optional Longitudinal Phase-Center Analysis}
\label{sec:asa-longitudinal}

ASA can also compare decoded phase organization across sessions.  In
a modeling dataset with sessions 0--9, grid cells were grouped into
modules and the same module was tracked across training.  This
  analysis is useful as a workflow demonstration because it combines
  module classification, TDA, cohomology decoding, EcohoSpace phase
  centers, and toroidal minimum-image displacement.  We treat this
  longitudinal analysis as a secondary extension of ASA rather than as
  a required result for the present main-text argument.

\subsection{Interpretation}
\label{sec:asa-interpretation}

The revised application structure reflects ASA's intended role.  ASA
is not a torus-only detector and should not be judged only by whether
a single barcode looks like $T^2$.  It is a workflow for testing
attractor topology at multiple levels: whether a population trajectory
contains one stable circular coordinate, whether a grid-cell module
contains two stable circular coordinates and a compatible $H_2$
feature, whether decoded cohomology coordinates map back to physical
or behavioural variables, whether individual cells concentrate in
cohomology space, and whether topology-aware cell selection improves
the result.  Real MEC modules are the main empirical testbed, while
virtual CANN/RNN datasets and spatial TDA serve as controlled
benchmarks and supplementary demonstrations.

%% file: sections/backmatter.tex

\section{Discussion and Conclusion}

The CANNs ecosystem presented in this report---one Python
library, one Rust acceleration backend, and one experimental-data
pipeline, all released under permissive licences and
co-developed by a single research group---is a deliberately
narrow answer to a deliberately broad problem.  We have not
attempted to compete with the generality of NEST or Brian~2, nor
to replace the analysis depth of gudhi or Ripser; we have
instead tried to package the small set of CANN-specific primitives
that a research group in this area reaches for again and again,
and to expose them through an API that does not punish the user
for not also being an expert in numerical analysis, persistent
homology, and Rust FFI.  The combination of an English and a
Chinese documentation set, a self-contained Docker image, a
single \texttt{make reproduce} command, and a graphical front
end for the analysis pipeline is the most significant deliverable of
this work; the remaining components are implemented in a few
hundred lines of Python.

\subsection{Limitations}

Several open items are genuine limitations of the present release:

\textbf{$H_2$-homology scalability.}  Computing second homology
($H_2$) over a Vietoris--Rips filtration is combinatorially more
expensive than $H_0$ or $H_1$, and the current Rust-backed Ripser
implementation does not yet benefit from the same parallelisation
optimisations as the $H_1$ path.  Users analysing very large
point clouds (~$n > 5{,}000$) for toroidal topology should expect
longer runtimes for the $H_2$ stage.

\textbf{Absence of a band-cell classifier.}  Band cells, which
encode a single spatial frequency along one direction, are an
important intermediate population between grid cells and place cells,
but ASA currently does not provide a dedicated classifier for band-like
topology.  Detecting band structures requires the same $H_1$ analysis
as ring attractors, but with an anisotropic distance metric; this is a
planned extension.

\textbf{Manual parameter workflow.}  ASA requires users to specify
the smoothing kernel width $\sigma$, the PCA truncation rank, the
filtration metric, and the shuffle threshold $q$.  While default
values are provided (see Sec~\ref{sec:asa}), there is currently no
automated recommendation engine that suggests parameters based on
data characteristics.  Users are expected to exercise scientific
judgment when selecting these values.

\subsection{Roadmap}

The project is actively developed and the issue trackers on both the
\pkg{canns} and \pkg{canns-lib} repositories track open work.  The
modular architecture means that external contributions can land without
disturbing the rest of the codebase.  Near-term planned features
include: (i) an accelerated $H_2$-filtration kernel in the Rust
backend; (ii) a band-cell classification module in ASA; (iii) an
approximate nearest-neighbour search for speeding up large-trajectory
navigation; (iv) a shared-memory parallel batch evaluator for
high-throughput screening; and (v) a parameter recommendation system
that suggests $\sigma$, PCA rank, and $q$ based on data statistics.
We invite interested readers to file issues, send pull requests, or
fork the toolkit to fit their own experimental questions.

\section{AI Usage Disclosure}

We used AI-assisted tools during the development of this work. Their
roles were limited to (i)~assisting with code review, refactoring, and
documentation of the \texttt{canns} and \texttt{canns-lib} codebases,
(ii)~language polishing and consistency checks of the documentation
strings, docstrings, and tutorial notebooks, and (iii)~suggesting
LaTeX structural templates and reference-management boilerplate used
in this paper. All core scientific code, mathematical derivations,
model implementations, and analysis pipelines were written by human
authors. All AI-generated text and code suggestions were reviewed,
edited where necessary, and validated against the underlying
scientific claims by the authors. The authors take full
responsibility for the content of this paper.

\section{Acknowledgments}

We gratefully acknowledge the experimental collaborators who provided
neural recordings used to validate the ASA pipeline.
This work was supported by the National Natural Science Foundation
of China (No.\ T2421004 to S.Wu), the National Key Research and
Development Program of China (No.\ 2024YFF1206500), and the Science
and Technology Innovation 2030 -- Brain Science and Brain-inspired
Intelligence Project (No.\ 2021ZD0200204 to S.Wu).